\documentclass[aps,prl,amsmath,amssymb,reprint,superscriptaddress]{revtex4-1}
\usepackage{braket}
\usepackage{mycommands}
\usepackage{hyperref}
\usepackage[capitalise,]{cleveref}
\usepackage{comment}
\usepackage{chemfig}

\usepackage{bm}

\usepackage{siunitx}
\usepackage{bbold}
\usepackage{dsfont}
\usepackage{amsmath}
\usepackage{mathtools}
\usepackage[T1]{fontenc}
\allowdisplaybreaks

\DeclareSIUnit\rydberg{Ry}

\graphicspath{{./figures/}}

\begin{document}

\title{A Scalable Translationally Invariant Variational Theory of Ab Initio Polarons} 

\author{Moritz K. A. Baumgarten}
\affiliation{Department of Chemistry and Chemical Biology, Harvard University, Cambridge, MA, USA}

\author{Hamlin Wu}
\affiliation{Department of Chemistry and Chemical Biology, Harvard University, Cambridge, MA, USA}

\author{Tong Jiang}
\affiliation{Department of Chemistry and Chemical Biology, Harvard University, Cambridge, MA, USA}

\author{Joonho Lee}
\email{joonholee@g.harvard.edu}
\affiliation{Department of Chemistry and Chemical Biology, Harvard University, Cambridge, MA, USA}

\date{\today}

\begin{abstract}
We introduce a scalable, translationally invariant variational theory for {\it ab initio} polarons that remains applicable across coupling regimes without resorting to supercells. Our approach combines a momentum-projected Toyozawa-type wavefunction with a low-rank factorization of the electron--phonon kernel, enabling near-linear scaling with the number of $\mathbf{k}$-points while capturing both delocalized and self-trapped carriers. Benchmarks for the Fr\"ohlich model, \ce{LiF}, and anatase and rutile \ce{TiO2} yield accurate polaron binding energies, thermodynamic-limit band structures, and transparent real-space measures of polaron extent. For \ce{LiF}, comparison with first-principles diagrammatic Monte Carlo (DiagMC) reveals close agreement for the weak-coupling electron-polaron ground state and band structure. 
However, in the hole-polaron of LiF, which is in the strong-coupling regime, we found a significant bias in DiagMC results.
These results establish momentum-projected variational wavefunctions as a systematically improvable route to thermodynamic limit studies of polarons in real materials.
\end{abstract}

\maketitle

{\it Introduction.} 
Polarons, charge carriers dressed by lattice vibrations, underlie a broad range of phenomena in condensed-matter systems~\cite{landau1933electron,giustino_electron-phonon_2017,bardeen_theory_1957,marsiglio2020eliashberg}. In realistic materials an all-coupling {\it ab initio} description that remains both translationally invariant and computationally tractable on dense Brillouin-zone meshes is still lacking. Perturbative and Green's-function-based methods~\cite{fan1951temperature, migdal1958interaction, allen_theory_1976, allen_theory_1981,story2014cumulant,lihm2025nonperturbative} are most effective when lattice dressing is modest, while existing variational formulations either favor the strong-coupling regime~\cite{sio_polarons_2019,sio_ab_2019,luo_comparison_2022} or become prohibitively expensive~\cite{mahajan_structure_2025,robinson_ab_2025} for thermodynamic-limit (TDL) extrapolations. Other notable recent developments include the self-consistent Green's-function method~\cite{lafuente-bartolome_ab_2022,lafuente-bartolome_unified_2022} and first-principles diagrammatic Monte Carlo (DiagMC)~\cite{luo_first-principles_2025}. DiagMC is, in principle, unbiased but becomes increasingly difficult to sample in strongly interacting regimes~\cite{carlstrom2021strong}. In this Letter, we introduce a scalable, translationally invariant variational framework for {\it ab initio} polarons that remains applicable across coupling regimes without supercells.

Our approach extends the momentum-conserving Toyozawa wavefunction~\cite{toyozawa_self-trapping_1961,zhao_hierarchy_2023,shih_theory_2024} to first-principles electron--phonon Hamiltonians and, through a low-rank factorization of the electron--phonon kernel~\cite{luo_data-driven_2024}, reduces the dominant computational bottleneck to near-linear scaling with Brillouin-zone sampling. We validate the resulting wavefunction for the Fr\"ohlich model, \ce{LiF}, and anatase and rutile \ce{TiO2}. Together these systems span weak-coupling, strong-coupling, and anisotropic large-polaron regimes relevant to current state-of-the-art first-principles calculations. The method yields accurate ground-state energies, TDL band structures, and straightforward real-space correlation functions, establishing a systematically improvable alternative to existing first-principles approaches for polaron problems.

{\it Theory}. We consider the general first-principles electron--phonon (eph) Hamiltonian:
\begin{multline} \label{eq:elph_kspace}
    \hat{H} = {\sum_{n\mathbf{k}} \varepsilon_{n\mathbf{k}} c_{n\mathbf{k}}^\dagger c_{n\mathbf{k}}}+  {\sum_{\nu\mathbf{q}} \omega_{\nu \mathbf{q}} b_{\nu \mathbf{q}}^\dagger b_{\nu \mathbf{q}}}\\ +{\frac{1}{\sqrt{N_\mathbf{k}}}\sum_{\substack{nm\nu \\ \mathbf{k} \mathbf{q}}} g_{mn\nu}(\mathbf{k},\mathbf{q}) c_{m\mathbf{k}+\mathbf{q}}^{\dagger} c_{n\mathbf{k}} (b_{\nu-\mathbf{q}}^\dagger+ b_{\nu\mathbf{q}})} 
\end{multline}
where $c_{n\mathbf{k}}$ and $b_{\nu \mathbf{q}}$ are electron and phonon annihilation operators, respectively, and $\varepsilon_{n\mathbf{k}}$, $g_{mn\nu}(\mathbf{k},\mathbf{q})$, and $\omega_{\nu \mathbf{q}}$ are the electronic bands, eph coupling matrix elements, and phonon frequencies, respectively.
In this setting, weak-coupling regimes can often be described perturbatively, whereas strong-coupling regimes are naturally captured by localized adiabatic states~\cite{franchini_polarons_2021}. A central challenge for {\it ab initio} polaron theory is therefore to construct a wavefunction that interpolates between these limits while preserving translational invariance (or constructing an eigenstate of the {\it total} momentum operator) and remaining computationally tractable on dense Brillouin-zone meshes.

\begin{figure}[!ht]
    \centering
    \includegraphics[width=0.98\linewidth]{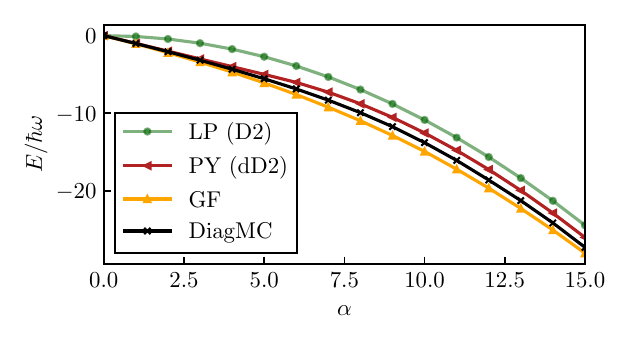}
    \caption{Ground state energy of the Fröhlich polaron for numerous methods as a function of the coupling strength $\alpha$. Exact DiagMC values were extracted from Ref.~\cite{hahn_diagrammatic_2018}. Self-consistent Green's function (GF) data were extracted from Ref.~\cite{lafuente-bartolome_unified_2022}. Numerical details of LP and PY are given in \cref{sec:frohlich}.}
    \label{fig:frohlich}
\end{figure}

Our starting point is the observation that strong-coupling polaron wavefunctions are naturally expressed in terms of localized electron--lattice distortions, whereas weak-coupling polarons require a translationally invariant description. We bridge these limits by applying momentum projection to an adiabatic localized state. Specifically, we begin from the Landau--Pekar (LP) product state~\cite{landau1948effective},
\begin{equation}
    |\Psi_{\text{LP}}(\mathbf r)\rangle = \psi_{\text{el}}(\mathbf r) \otimes |f\rangle
\end{equation}
where $\psi_{\text{el}}(\mathbf r)$ is a localized electron state (often modeled as a Gaussian), and $|f\rangle = \exp(\sum_{\mathbf{q}} f_{\mathbf{q}} b_{\mathbf{q}}^\dagger - f_{\mathbf{q}}^* b_{\mathbf{q}})|0\rangle$ is a coherent state representing the lattice deformation. 
We then restore translational symmetry through the Peierls--Yoccoz (PY) projection operator~\cite{peierls1957collective}, constructing a momentum eigenstate that retains the localized strong-coupling physics of the adiabatic state while recovering the delocalized character required in the weak-coupling limit.
The resulting effectiveness of momentum-projection is shown in \cref{fig:frohlich}.

We obtain a variational energy that recovers the correct weak-coupling ($E \propto \alpha$) and strong-coupling ($E \propto \alpha^2$) limits of the Fr\"ohlich model, while substantially improving over the unprojected LP state. Our results are comparable to those of the all-coupling Green's function method by Giustino and co-workers~\cite{lafuente-bartolome_ab_2022,lafuente-bartolome_unified_2022} and the {\it ab initio} Nagy--Markos method studied by Reichman and co-workers~\cite{nagy_nature_1989,robinson_ab_2025}. Earlier work by Pekar had already emphasized the importance of restoring translational invariance in Fr\"ohlich-type polarons~\cite{buimistrov_quantum_1958}. In this Letter, we show that the same idea can be implemented variationally and extended to realistic {\it ab initio} Hamiltonians.

In a finite Bloch and phonon basis, the Landau--Pekar construction leads to the second-kind Davydov (D2) wavefunction,
\begin{equation} \label{eq:d2}
    |\Psi_{\mathrm{D}2}\rangle = \sum_{n \mathbf{k}} A_{n\mathbf{k}} c_{n\mathbf{k}}^\dagger |0\rangle \otimes \bigotimes_{\nu \mathbf{q}} \hat{\mathcal{D}}(B_{\nu \mathbf{q}})|0\rangle,
\end{equation}
where $A_{n\mathbf{k}}$ and $B_{\nu \mathbf{q}}$ are variational parameters. The displacement operator $\hat{\mathcal{D}}(B_{\nu \mathbf{q}}) = \exp(B_{\nu \mathbf{q}}b_{\nu \mathbf{q}}^\dagger - B_{\nu \mathbf{q}}^*b_{\nu \mathbf{q}})$ generates a coherent state of mode $(\nu,\mathbf{q})$ with amplitude $B_{\nu \mathbf{q}}$. The D2 ansatz was originally introduced by Davydov~\cite{davydov_theory_1973} and later adapted to {\it ab initio} polaron problems by Sio and co-workers~\cite{sio_ab_2019,luo_comparison_2022}. Variational minimization of D2 scales as $\mathcal{O}(N_\mathbf{k}^2 N_b^2 N_{\text{mod}})$, where $N_\mathbf{k}$ is the number of $\mathbf{k}$-points, $N_b$ the number of electronic bands, and $N_{\text{mod}}$ the number of phonon modes. 
A perturbative correction on top of D2 states, coherent-state second-order perturbation theory (CSPT2)~\cite{robinson_ab_2025}, can improve them. CSPT2 is especially accurate in low-phonon excitation regimes of D2 with a higher cost of $\mathcal{O}(N_{\mathbf{k}}^3 N_b^2 \max\{N_b, N_{\text{mod}}\})$.

Because Eq.~\eqref{eq:d2} explicitly breaks translational symmetry, it becomes inadequate in regimes where delocalization of the excess charge carrier is essential. We therefore project Eq.~\eqref{eq:d2} onto a crystal-momentum sector labeled by $\mathbf{K}$ using the discretized PY projector. This yields the delocalized D2 (dD2) wavefunction,
\begin{align} \label{eq:dd2}
    |\Psi_{\mathrm{dD2}}^\mathbf{K}\rangle &= \sum_{j} \mathrm{e}^{\mathrm{i}(\mathbf{K} - \hat{K})\cdot\mathbf{R}_j} |\Psi_{\mathrm{D}2}\rangle,
\end{align}
where $\hat{K} = \sum_{n\mathbf{k}} \mathbf{k}\, c_{n\mathbf{k}}^\dagger c_{n\mathbf{k}} + \sum_{\nu\mathbf{q}} \mathbf{q}\, b_{\nu \mathbf{q}}^\dagger b_{\nu \mathbf{q}}$ is the total crystal-momentum operator and $\mathbf{R}_j$ runs over lattice translations within the Born--von Karman supercell. This projected state is the {\it ab initio} counterpart of the Toyozawa wavefunction~\cite{toyozawa_self-trapping_1961}. Related constructions have been studied extensively for model Hamiltonians~\cite{zhao_variational_1997,shih_theory_2024}, and have recently served as a reference wavefunction for variational Monte Carlo (VMC) calculations~\cite{mahajan_structure_2025}. However, their direct application to realistic {\it ab initio} systems has been hindered by unfavorable scaling with Brillouin-zone sampling.

To expose the bottleneck, we examine the variational energy of the dD2 wavefunction. The gradient evaluation follows a similar contraction path and therefore has the same asymptotic scaling as the energy evaluation~\cite{griewank2014automatic}. Up to a normalization factor, the variational  energy is given by
\begin{align} \label{eq:dd2_var_energy}
\begin{split}
     &E_\mathrm{dD2} (\mathbf K) = 
     \sum_{n\mathbf{k}} |A_{n\mathbf{k}}|^2 \left(\varepsilon_{n\mathbf{k}}D_\mathbf{k} 
    + \sum_{\nu \mathbf{q}} \omega_{\nu \mathbf{q}}|B_{\nu \mathbf{q}}|^2 D_{\mathbf{k}+\mathbf{q}}\right)\\
    &+ \frac{2}{\sqrt{N_k}}\:\operatorname{Re}\bigg\{\sum_{mn\nu\mathbf{kq}} g_{mn\nu}(\mathbf{k},\mathbf{q}) A^*_{m\mathbf{k}+\mathbf{q}} A_{n\mathbf{k}}B_{\nu-\mathbf{q}}^{*} D_{\mathbf{k}} \bigg\},
\end{split}
\end{align}
where we used the hermiticity of the eph coupling matrix elements and introduced the intermediate,
$
    D_{\mathbf{k}} = \sum_j \mathrm{e}^{\mathrm{i}\left(\mathbf{K}- \mathbf{k}\right) \cdot\mathbf{R}_j} \: \mathrm{e}^{\sum_{\nu\mathbf{q}}|B_{\nu\mathbf{q}}|^2\mathrm{e}^{-\mathrm{i}\mathbf{q}\cdot \mathbf{R}_j}}.
$
The dominant cost arises from the eph contribution, namely the last term in Eq.~\eqref{eq:dd2_var_energy}, which contains coupled summations over $\mathbf{k}$ and $\mathbf{q}$ and therefore scales as $\mathcal{O}(N_{\mathbf{k}}^2 N_{\mathrm{b}}^2 N_{\mathrm{mod}})$, same as that of D2. This quadratic scaling in $N_{\mathbf{k}}$ makes dense Brillouin-zone sampling prohibitively expensive, especially in the large-polaron regime, where reliable extrapolation to the TDL requires very fine meshes. Our key contribution in this Letter is to show that this contraction can be reorganized into uncoupled $\mathbf{k}$ and $\mathbf{q}$ summations when the eph kernel is expressed in low-rank form.

\begin{figure*}[!ht]
    \centering
    \includegraphics[width=\linewidth]{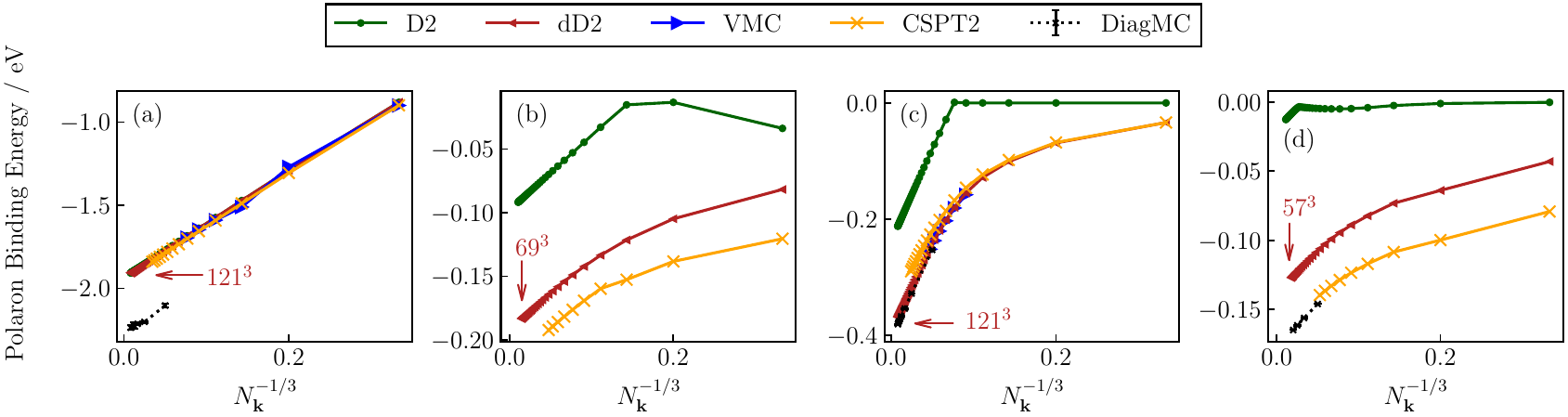}
    \caption{Size extrapolation for the small (a) hole polaron in \ce{LiF} and (b) electron polaron in rutile, and large electron polarons in (c) \ce{LiF} and  (d) anatase. We indicated the largest $\mathbf{k}$-meshes run for each system with dD2 in each panel. VMC values were taken from Ref.~\cite{mahajan_structure_2025}. DiagMC values were taken from Ref.~\cite{luo_first-principles_2025}. D2, dD2, and CSPT2 calculations were performed retaining singular values corresponding to a relative error in $g_{mn\nu}(\mathbf{k},\mathbf{q})$ of $10^{-3}$. A detailed analysis of the SVD truncation error is provided in \cref{sec:svd_error}.
    }
    \label{fig:polaron_finite_size_gs}
\end{figure*}

We adopt a recently introduced low-rank representation of the short-range contribution to the eph matrix elements~\cite{luo_data-driven_2024},
\begin{align} \label{eq:g_svd}
    \begin{split}
        g_{mn\nu}(\mathbf{k},\mathbf{q}) &=
        \sum_{i} U_{im}^*(\mathbf{k}+\mathbf{q})\, U_{in}(\mathbf{k}) L_{\nu}(\mathbf{q}) \\
        &\quad + \sum_{ij\gamma} U_{im}^*(\mathbf{k}+\mathbf{q}) \, \Sigma_{ij}^\gamma(\mathbf{k}) \, V^\gamma_{ij\nu}(\mathbf{q}) \, U_{jn}(\mathbf{k}),
    \end{split}
\end{align}
where $i,j$ are Wannier orbital indices, $\gamma$ indexes the $N_c$ retained singular vectors, $U_{im}(\mathbf{k})$ is the Wannier-to-Bloch transformation matrix on the dense grid interpolated from a coarse grid~\cite{giustino2007electron}, $L_\nu(\mathbf{q})$ is the analytical long-range contribution from the multipole expansion~\cite{vogl_microscopic_1976,verdi_frohlich_2015,park_long-range_2020}, and $\Sigma_{ij}^\gamma(\mathbf{k})$ and $V_{ij\nu}^\gamma(\mathbf{q})$ are singular vectors. More details on the computation of $g_{mn\nu}(\mathbf{k},\mathbf{q})$ are provided in ~\cref{sec:elph_coupling_elements}. 
While Eq.~\eqref{eq:g_svd} was introduced to reduce the storage costs for {\it ab initio} eph kernels, here we show that it also reduces the computational cost by enabling a more favorable contraction order.

Substituting Eq.~\eqref{eq:g_svd} into the last term of Eq.~\eqref{eq:dd2_var_energy} (see \cref{sec:contract} for more details), 
the contraction over
$\mathbf{k}$ can now be evaluated in $\mathcal{O}(N_c N_\mathbf{k}\log N_\mathbf{k})$ time using the fast Fourier transform, a substantial improvement over the original $\mathcal{O}(N_\mathbf{k}^2)$ scaling. In practice, $N_c \sim 10^2-10^3$, whereas dense {\it ab initio} calculations require up to $N_\mathbf{k} \sim 10^5-10^6$. 
The low-rank decomposition in Eq.~\eqref{eq:g_svd} therefore provides two key benefits: it reduces the memory required to represent $g_{mn\nu}(\mathbf{k},\mathbf{q})$ on dense meshes, and it yields a near-linear scaling contraction path for variational energy evaluation. This intuition of computational scaling reduction applies to other methods than dD2, which we summarize in \cref{tab:complexity}.

\begin{table}[h]
    \centering
    \begin{tabular*}{\columnwidth}{@{\extracolsep{\fill}}lcc@{}}
    \hline
    \hline
      & No SVD & SVD \\ [0.8ex]
    \hline
    D2 &
    $\mathcal{O}\!\left(N_{\mathbf{k}}^2 N_b^2 N_{\text{mod}}\right)$ &
    $\mathcal{O}\!\left(N_c N_{\mathbf{k}} \log N_{\mathbf{k}}\, N_b^2 N_{\text{mod}}\right)$ \\
    
    dD2 &
    $\mathcal{O}\!\left(N_{\mathbf{k}}^2 N_b^2 N_{\text{mod}}\right)$ &
    $\mathcal{O}\!\left(N_c N_{\mathbf{k}} \log N_{\mathbf{k}}\, N_b^2 N_{\text{mod}}\right)$ \\
    
    CSPT2 &
    $\mathcal{O}\!\left(N_{\mathbf{k}}^3 N_b^2 \max\{N_b, N_{\text{mod}}\}\right)$ &
    $\mathcal{O}\!\left(
    \begin{array}{c}
    N_b^2 N_{\mathbf{k}}^2 \max\!\left(N_{\mathbf{k}} N_b,\right.\\
    \left. N_c \log N_{\mathbf{k}}\, N_{\text{mod}}\right)
    \end{array}
    \right)$ \\
    \hline
    \hline
    \end{tabular*}
    \caption{Time complexity associated with the methods considered in this work.}
    \label{tab:complexity}
\end{table}

{\it Polaron binding energy}. To assess the method in a first-principles setting, we consider the binding energies of electron and hole polarons in \ce{LiF} and of electron polarons in anatase and rutile \ce{TiO2}. In all cases, reliable comparison requires extrapolation to the TDL using a linear $1/L$ form~\cite{makov_periodic_1995}, as shown in \cref{fig:polaron_finite_size_gs}. We performed calculations on meshes up to $121^3$ for \ce{LiF}, $57^3$ for anatase, and $69^3$ for rutile. The extrapolated binding energies are summarized in Table~\ref{tab:formation_energies_P}; additional details of the finite-size analysis are given in \cref{sec:SI_size_extra}.

\begin{figure*}[!ht]
    \centering
    \includegraphics[width=\linewidth]{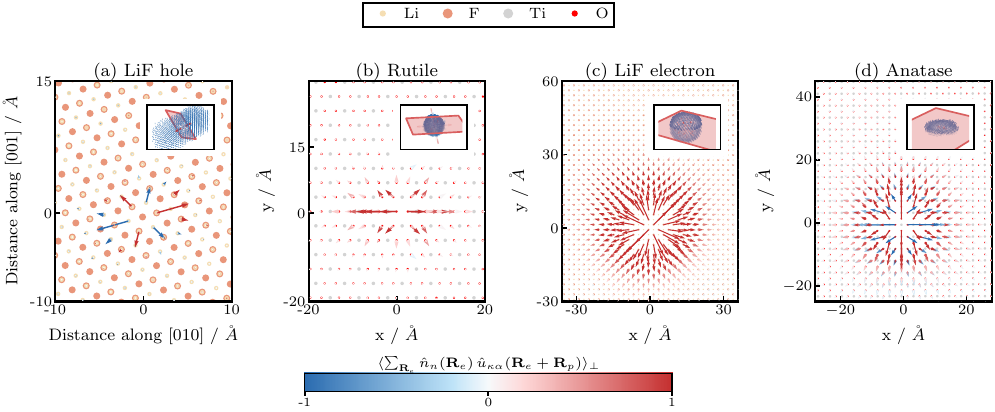}
    \caption{Visualization of the density--displacement correlator for (a) the \ce{LiF} hole polaron along the [010]-[001] plane, and for (b) the rutile \ce{TiO2} electron polaron, (c) the \ce{LiF} electron polaron, and (d) the anatase \ce{TiO2} electron polaron, sliced across the $xy$-plane. Arrows indicate the directed correlation and have been scaled by factors $6\times$, $2000\times$, $1000\times$, and $366\times$ in panels (a)-(d), respectively. The colorbar encodes a normalized in- and out-of-plane amplitude of the correlator. Insets show $0.05\%$ of elements with the largest magnitudes.}
    \label{fig:extent}
\end{figure*}

In \cref{fig:polaron_finite_size_gs}(a), we show the size scaling for the \ce{LiF} hole polaron, a prototypical strong-coupling reference system~\cite{robinson_ab_2025,luo_first-principles_2025,mahajan_structure_2025,sio_ab_2019}. In this case, the dD2 solution collapses onto the D2 solution, yielding a binding energy of $1.933$~\si{\electronvolt}. This is expected when the polaron is sufficiently localized. Overlaps between translated D2 configurations become negligible, and hence symmetry projection only yields a minor improvement in the ground state energy. The resulting energies are in excellent agreement with VMC, and CSPT2 adds only a small $17$~\si{\milli\electronvolt} correction, further supporting the accuracy of D2 and dD2 in this regime. We note that the small energy difference between D2 and CSPT2 suggests that the eph correlation beyond D2 is negligible, making D2 effectively near-exact for this system. By contrast, the published first-principles DiagMC value is substantially larger at $2.26$~\si{\electronvolt}~\cite{luo_first-principles_2025}. The agreement among D2, dD2, CSPT2, and VMC, together with the much larger binding energy of DiagMC, indicates systematic error in the published first-principles DiagMC result for this strong-coupling LiF hole-polaron case.

In \cref{fig:polaron_finite_size_gs}(b), we consider the electron polaron in rutile \ce{TiO2}. Here, D2 develops a localized solution already on a relatively coarse $7^3$ $\mathbf{k}$-mesh, while momentum projection on top lowers the energy substantially further. 
At the same time, the phonon statistics remain concentrated in low phonon-number sectors, and the electronic density is strongly centered at $\Gamma$, indicating a relatively weak dressing of the polaron ground state. 
In this low-phonon regime, CSPT2 built on the undisplaced D2 reference yields the lowest energy among the approaches considered here.

For the \ce{LiF} electron polaron in \cref{fig:polaron_finite_size_gs}(c), the momentum-projected dD2 wavefunction yields a binding energy of $-0.395$~\si{\electronvolt}, in close agreement with the extrapolated DiagMC value of $-0.408$~\si{\electronvolt} and with VMC results~\cite{mahajan_structure_2025}. In contrast, D2 substantially underbinds because it cannot capture the delocalized character of the polaron in the weak-to-intermediate-coupling regime. We also report CSPT2 results based on the undisplaced D2 reference~\cite{lee_constrained-path_2021,robinson_ab_2025}. In this case, one-phonon sectors account for only a small fraction of the ground-state weight, so CSPT2 remains less accurate than dD2. Correspondingly, dD2 yields a broad phonon-number distribution, consistent with DiagMC, as shown in \cref{subsec:phonon_number}.

Finally, for anatase \ce{TiO2} in \cref{fig:polaron_finite_size_gs}(d), D2 gives only a small binding energy of $20$~\si{\milli\electronvolt}, comparable to the $10$~\si{\milli\electronvolt} reported by Dai and Giustino~\cite{dai_identification_2024}. In contrast, dD2 stabilizes the polaron substantially, giving a binding energy of $-0.138$~\si{\electronvolt}. The dominant weight in low phonon-number sectors indicates weak lattice dressing and a largely Bloch-like carrier, so restoring translational invariance is essential for an accurate ground-state description. In this regime, CSPT2 performs better than dD2 because its first-order wavefunction explicitly spans the $0$- and $1$-phonon sectors coupled to the electron. Nevertheless, this improved accuracy comes at a significantly higher computational cost, as summarized in \cref{tab:complexity}. To further characterize the dD2 states obtained here, we provide carrier occupations in \cref{subsec:elec_obs}.

\begin{table}[h]
    \centering
    \begin{tabular*}{\columnwidth}{@{\extracolsep{\fill}}lcccc@{}}
     \hline
     \hline
      & D2 & dD2 & CSPT2 & DiagMC \\ [0.8ex]
     \hline
     LiF (e) & --0.239  & --0.395  &  --0.363 & --0.4084(22)  \\
     LiF (h) & 1.933  & 1.933 & 1.950 & 2.260(29) \\
     Anatase & --0.020 & --0.138 & --0.165  & --0.17798(13) \\
     Rutile & --0.098 & --0.192 & --0.219  & -- \\ [0.5ex]
     \hline
     \hline
    \end{tabular*}
    \caption{Extrapolated polaron formation energies in \si{\electronvolt}. DiagMC values are taken from Ref.~\cite{luo_first-principles_2025}. We provide more details on the size extrapolation in \cref{sec:SI_size_extra}.
    }
    \label{tab:formation_energies_P}
\end{table}

{\it Polaron extent.} Variational wavefunctions also provide direct access to observables beyond the ground-state energy. In particular, they allow us to characterize the spatial extent of the polaron. 
Because dD2 is translationally invariant, spatial extent must be inferred from a two-point correlation function, for which we use the density--displacement correlator,
\begin{align} \label{eq:extent}
    \eta_{n\kappa\alpha}(\mathbf{R}_p)
    &= \left\langle \sum_{\mathbf{R}_e} \hat{n}_n(\mathbf{R}_e)\,
    \hat{u}_{\kappa\alpha}(\mathbf{R}_e+\mathbf{R}_p)\right\rangle, 
\end{align}
where $\kappa$, $\alpha$, and $n$ denote the atom,  the displacement direction, and the Wannier orbital, respectively, $\hat{n}_n(\mathbf{R})$ is the density operator of the $n$-th Wannier orbital in the $\mathbf R$-th cell and $\hat{u}_{\kappa\alpha}(\mathbf{R})$ is the displacement operator of the $\alpha$-th component of the $\kappa$-th atom in the $\mathbf R$-th cell. 
This correlator has long been used as a proxy for polaron size~\cite{scherer_theory_1984,romero_exact_1999,hoffmann_optical_2002}, and was recently applied to the anatase electron polaron using DiagMC~\cite{liu_understanding_2026}. Further details of its evaluation are given in \cref{sec:polaron_extent}.

In \cref{fig:extent}, we visualize this correlator for the four systems studied here. This analysis shows that the electron polarons in \ce{LiF} and anatase are substantially more extended than the electron polaron in rutile and the hole polaron in \ce{LiF}. The \ce{LiF} electron polaron is nearly isotropic, whereas the anatase electron polaron is strongly anisotropic and predominantly two-dimensional, consistent with previous work~\cite{dai_identification_2024,liu_understanding_2026}. Rutile is more compact and anisotropic, while the hole polaron in \ce{LiF} is the most localized of the four cases.

\begin{figure}[!ht]
    \centering
    \includegraphics[width=\linewidth]{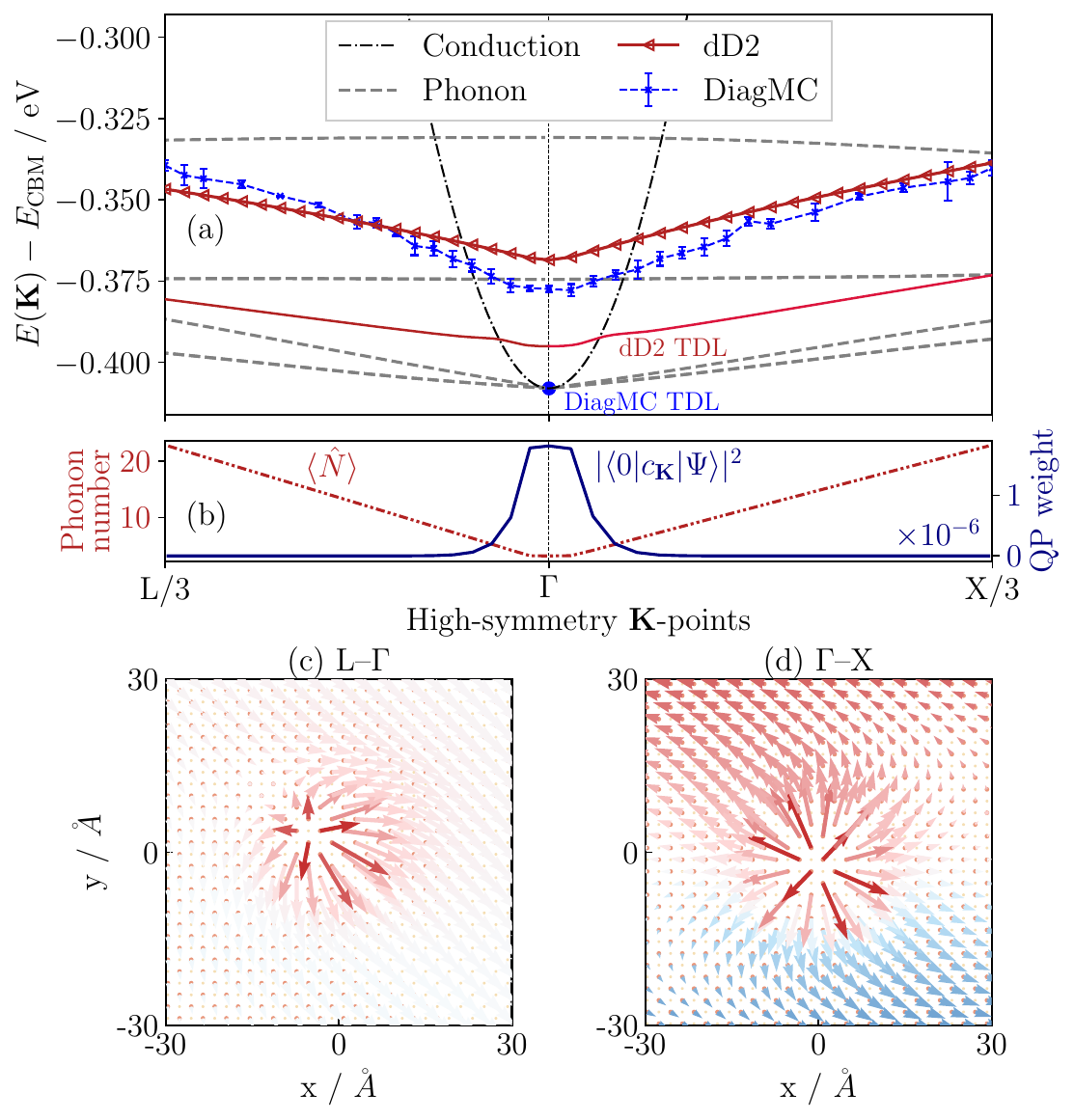}
 \caption{(a) Band structure of the \ce{LiF} electron polaron. The conduction and phonon bands are shifted to the extrapolated DiagMC ground-state energy at $\mathbf{K}=\Gamma$. (b) Quasiparticle weight $|\langle0|c_{\mathbf{K}}|\Psi_{\text{dD2}}\rangle|^2$ and phonon number $\langle \hat{N}\rangle$ across $\mathbf{K}$. Extent of the LiF electron polaron for (c) $\mathbf{K} = (\frac{31}{121}, \frac{31}{121}, \frac{31}{121})$ and (d) $\mathbf{K} = (\frac{31}{121}, 0, \frac{31}{121})$.
 }
    \label{fig:bandstructure}
\end{figure}

{\it Polaron band structures}. Because dD2 explicitly conserves the total crystal momentum $\mathbf{K}$, the variational optimization can be carried out independently in each momentum sector, giving direct access to the polaron dispersion. In \cref{fig:bandstructure}, we present the band structure of the \ce{LiF} electron polaron obtained on a $120^3$ $\mathbf{K}$ mesh and extrapolated to the TDL. This enables a thermodynamic-limit extrapolation of an {\it ab initio} polaron band using dD2.
Near $\Gamma$, the dD2 dispersion is in excellent agreement with first-principles DiagMC, reproducing the overall bandwidth and effective masses. 

Away from $\Gamma$, along the L--$\Gamma$ line roughly between L/3 and L/6, the finite-mesh dD2 energies lie systematically below the corresponding first-principles DiagMC values on the same $\mathbf{K}$ mesh. 
Because dD2 is variational, this establishes residual error in the published DiagMC energies in these momentum sectors. 
In several momenta the discrepancy also exceeds the reported DiagMC statistical error bars, indicating that stochastic uncertainty alone is unlikely to account for the difference. 
We trace this discrepancy to inaccuracies in the eph Hamiltonian used in the original DiagMC study, as detailed in \cref{sec:bandbias}. 
Furthermore, the phonon number increases away from $\Gamma$, indicating stronger effective dressing precisely where the DiagMC error bars grow. 
This is consistent with systematic sampling difficulty in these more strongly coupled systems, much as inferred above for the LiF hole polaron.

Finally, to illustrate how the polaron's real-space structure evolves away from the band minimum, panels (c) and (d) of \cref{fig:bandstructure} show the density--displacement correlation at two finite crystal momenta. At $\Gamma$, the polaron preserves the full symmetry prescribed by the corresponding matrix elements. At finite $\mathbf{K}$, the correlation develops a pronounced directional character along the propagation axis: a visible tail along [111] at L--$\Gamma$, and a similar bias along $y$ at $\Gamma$--X. The latter is accompanied by a faint, long-wavelength Bloch modulation, visible as the smooth red-to-blue gradient across the slice. These directional features are a real-space manifestation of the finite group velocity that the polaron acquires away from $\Gamma$.

{\it Conclusions.} 
We have developed a scalable, translationally invariant variational theory for {\it ab initio} polarons that unifies weak-, intermediate-, and strong-coupling regimes without resorting to supercells. 
The central ingredients are a momentum-projected Toyozawa-type wavefunction and a low-rank factorization of the electron--phonon kernel, which together enable near-linear scaling with Brillouin-zone sampling while capturing both delocalized and self-trapped carriers.
Across the Fr\"ohlich model, \ce{LiF}, and anatase and rutile \ce{TiO2}, the method delivers accurate polaron binding energies, TDL band structures, and transparent real-space characterization of polaron extent. We further highlight a strong-coupling challenge for first-principles DiagMC, as exemplified by the LiF hole polaron. Since dD2 is the simplest mean-field wavefunction that is translationally invariant, we expect that a systematically improvable wavefunction hierarchy, such as perturbation theory, can be developed on top of dD2. Our work establishes momentum-projected variational wavefunctions as a powerful foundation for predictive, TDL studies of polarons in real materials.

{\it Computational details}. We implemented {\it ab initio} Davydov methodologies in a development version of \texttt{Q-Chem}~\cite{epifanovsky_software_2021} and performed variational optimization using geometry-direct minimization~\cite{van_voorhis_geometric_2002}. 
All Hamiltonian matrix elements were obtained with \texttt{Quantum Espresso}~\cite{giannozzi_quantum_2009}, and a modified version of \texttt{Perturbo}~\cite{zhou_perturbo_2021}, and \texttt{EPW}~\cite{ponce_epw_2016}. More details are provided in \cref{sec:derivatives} and \cref{sec:comp_details}.

{\it Data availability}. \texttt{Quantum Espresso} inputs, \texttt{Perturbo} inputs, and source data to reproduce plots presented in this work are available under \url{https://github.com/JoonhoLee-Group/ab_initio_dd2_data}.

{\it Acknowledgments.} This work was supported by Harvard University’s startup funds. 
T.J. also acknowledges support from the Gordon and Betty Moore Foundation Fellowship.
This work used computational resources from FASRC supported by the FAS Division of Science Research Computing Group at Harvard, and the Delta system at the National Center for Supercomputing Applications through allocation CHE250068 from the Advanced Cyberinfrastructure Coordination Ecosystem: Services \& Support (ACCESS) program, supported by National Science Foundation grants \#2138259, \#2138286, \#2138307, \#2137603, and \#2138296. The authors would like to thank Paul Robinson for helpful discussions. 

\bibliography{references_apr29, ref2}

\clearpage
\onecolumngrid

\appendix
\setcounter{secnumdepth}{2} 

\renewcommand{\thesection}{\Roman{section}}

\makeatletter
\@addtoreset{equation}{section}
\@addtoreset{figure}{section}
\@addtoreset{table}{section}
\makeatother

\renewcommand{\theequation}{\Alph{section}\arabic{equation}}
\renewcommand{\thefigure}{\Alph{section}\arabic{figure}}
\renewcommand{\thetable}{\Alph{section}\arabic{table}}

\section{Fröhlich model}\label{sec:frohlich}
To stress-test the dD2 wavefunction in the weak-coupling regime, we turn to the Fröhlich model,
\begin{align}
    \hat{H} = \frac{\hat{p}}{2} + \omega_{\text{LO}}\sum_{\mathbf{q}} b^\dagger_\mathbf{q} b_\mathbf{q} + \sum_{\mathbf{k}} V_{\mathbf{k}} b_\mathbf{k} \mathrm{e}^{\mathrm{i}\mathbf{k}\cdot \hat{\mathbf{r}}} + \text{h.c.}.
\end{align}
This can be discretized into a lattice model via insertions of resolutions of identity $\sum_{\mathbf{k}}|\mathbf{k}\rangle\langle\mathbf{k}|$ left and right, and after a rotation of the phonon creation and annihilation operators we arrive at
\begin{align}
    \hat{H} = \sum_{\mathbf{k}} \frac{\mathbf{k}^2}{2} c_{\mathbf{k}}^\dagger c_{\mathbf{k}} + \omega_{\text{LO}}\sum_{\mathbf{q}} b^\dagger_\mathbf{q} b_\mathbf{q} + \sum_{\mathbf{k}\mathbf{q}} V_{\mathbf{q}} c_{\mathbf{k}+\mathbf{q}}^\dagger c_{\mathbf{k}} \left(b^\dagger_{-\mathbf{q}} + b_{\mathbf{q}} \right).
\end{align}
Matrix elements $V_{\mathbf{q}}$ are given by
\begin{equation}
    V_{\mathbf{q}} = \left(\frac{2 \sqrt{2}\pi\alpha}{V}\right)^{\frac{1}{2}} \frac{1}{|\mathbf{q}|}
\end{equation}

\subsection{The Broken Symmetry of Strong Coupling}
In the strong-coupling regime ($\alpha \gg 1$), the standard variational approach is the Landau--Pekar (LP) product ansatz:
\begin{equation}
    |\Psi_{\text{LP}}\rangle = |\psi_e\rangle \otimes |f\rangle
\end{equation}
where $|\psi_e\rangle$ is a localized electron state (often modeled as a Gaussian, $\psi_e(\mathbf{r}) \propto \mathrm{e}^{-\frac{1}{2}\mu r^2}$), and $|f\rangle = \exp(\sum_{\mathbf{q}} f_{\mathbf{q}} a_{\mathbf{q}}^\dagger - f_{\mathbf{q}}^* a_{\mathbf{q}})|0\rangle$ is a coherent state representing the lattice deformation. 

By localizing the electron at a specific origin, the LP state breaks the translational invariance of the underlying Hamiltonian. It is not an eigenstate of the total momentum $\mathbf{P}_{\text{tot}}$. This results in an artificially high energy in the weak- and intermediate-coupling regimes.

\subsection{The Peierls--Yoccoz Projection}
To restore translational symmetry, we employ the Peierls--Yoccoz (PY) projection operator. Since the Hamiltonian is translationally invariant, states localized at different positions $\mathbf{R}$ must be degenerate. We construct a momentum eigenstate by integrating over all continuous spatial translations of the localized LP wavepacket:
\begin{equation}
    |\Psi_{\text{PY}}(\mathbf{P})\rangle = \int d^3\mathbf{R} \, \mathrm{e}^{\mathrm{i}\mathbf{P}\cdot\mathbf{R}} T(\mathbf{R}) |\Psi_{\text{LP}}\rangle
\end{equation}
where $T(\mathbf{R}) = \exp(-\mathrm{i} \mathbf{P}_{\text{tot}} \cdot \mathbf{R})$ is the translation operator. For the ground state, we project onto zero total momentum ($\mathbf{P} = 0$), reducing the ansatz to a uniform superposition of all displaced localized states:
\begin{equation}
    |\Psi_{\text{PY}}(0)\rangle = \int d^3\mathbf{R} \, |\Psi_{\text{LP}}(\mathbf{R})\rangle
\end{equation}
This projection effectively delocalizes the heavy polaron, thereby restoring the center-of-mass kinetic energy of the system.

\subsection{Variation After Projection (VAP)}
To correctly capture the crossover regime ($\alpha \approx 6$), one must use the Variation After Projection (VAP) method, as opposed to the projection after variation method. 
In VAP, the exact expectation value of the Hamiltonian is evaluated using the fully projected state:
\begin{equation}
    E_{\text{PY}} = \frac{\langle \Psi_{\text{PY}}(0) | H | \Psi_{\text{PY}}(0) \rangle}{\langle \Psi_{\text{PY}}(0) | \Psi_{\text{PY}}(0) \rangle}
\end{equation}
This yields a multidimensional integral over the translation coordinate $\mathbf{R}$. The variational parameters, such as the electron Gaussian width(s) $\mu$ and the phonon displacements $f_{\mathbf{q}}$ (parameterized by a recoil constant $\beta$), are optimized to minimize $E_{\text{PY}}$ directly. 

Because the optimizer has access to the fully projected energy landscape, it naturally discovers two limits:
\begin{itemize}
    \item Weak Coupling ($\alpha < 6$): The variation drives the localized width $\mu \to \infty$. The PY operator projects this localized singularity into an extended plane wave, smoothly recovering the Lee-Low-Pines limit ($E \approx -\alpha$).
    \item Strong Coupling ($\alpha > 6$): The variation settles on a finite $\mu$, reflecting the deep self-trapping potential, recovering the standard LP quadratic scaling ($E \propto -\alpha^2$).
\end{itemize}

By using a multi-Gaussian expansion (i.e., $N=3$) of the electron density alongside VAP, the Peierls-Yoccoz ansatz provides a highly accurate description of the Fröhlich polaron across all coupling regimes, without artificial discontinuities.
To evaluate the PY variational energy efficiently, the multi-dimensional integrations over electronic coordinates ($\mathbf{r}$) and phonon modes ($\mathbf{q}$) must be reduced. We employ a hybrid analytical-numerical approach, leveraging the properties of the 3-Gaussian ($N=3$) trial wavefunction to evaluate core matrix elements analytically, leaving only stable 1D integrals for numerical quadrature.

\subsection{Analytic Matrix Elements}
The localized electronic trial state is constructed as a linear combination of Gaussians, $\phi(\mathbf{r}) = \sum_{i=1}^3 c_i \mathrm{e}^{-\mu_i r^2 / 2}$. Because the generator coordinate shift $R$ acts as a simple translation, the overlap and kinetic energy matrix elements between the localized state and its shifted counterpart $\phi_R(\mathbf{r}) = \phi(\mathbf{r} - \mathbf{R})$ can be evaluated completely analytically. 

Defining the pair parameters $\mu_{ij} = \mu_i + \mu_j$ and the reduced parameter $\tilde{\mu}_{ij} = (\mu_i \mu_j)/\mu_{ij}$, the electronic overlap matrix elements in real space are:
\begin{equation}
    I_{ij}(R) = S_{ij} \exp\left(-\frac{1}{2} \tilde{\mu}_{ij} R^2 \right),
\end{equation}
where $S_{ij} = (2\pi / \mu_{ij})^{3/2}$ is the standard Gaussian overlap volume. The corresponding kinetic energy transfer matrix elements are analytically found to be:
\begin{equation}
    T_{ij}(R) = I_{ij}(R) \tilde{\mu}_{ij} \left( \frac{3}{2} - \frac{1}{2} \tilde{\mu}_{ij} R^2 \right).
\end{equation}
Furthermore, the Fourier transform of the electronic density, $\rho_q$, is evaluated analytically using the same Gaussian contraction rules, mapping the 3D density into a purely 1D radial function of momentum $q$,
\begin{align}
    \rho_q &= \langle \phi | \mathrm{e}^{\mathrm{i}\mathbf{q}\cdot\mathbf{r}} | \phi \rangle = \int d^3r \, |\phi(\mathbf{r})|^2 \mathrm{e}^{\mathrm{i}\mathbf{q}\cdot\mathbf{r}} \\
    &= \frac{1}{\mathcal{N}} \sum_{i=1}^3 \sum_{j=1}^3 c_i c_j S_{ij} \exp\left( -\frac{q^2}{2\mu_{ij}} \right),
\end{align}
where $\mathcal{N} = \sum_{i,j} c_i c_j S_{ij}$ is the normalization constant of the electronic state, and $\mu_{ij} = \mu_i + \mu_j$. These analytic integrations eliminate the need for spatial 3D quadrature, restricting the numerical cost purely to one-dimensional quadratures.

\subsection{Numerical Quadrature for Phonon and Interaction Matrix Elements}
The localized trial state assumes the lattice relaxes into a coherent state characterized by the displacement amplitudes $f_{\mathbf{q}}$. This displacement is phenomenologically tied to the exact electronic density via a single momentum-space low-pass filter:
\begin{equation}
    f_{\mathbf{q}} = - \frac{V_{\mathbf{q}}}{\hbar \omega_{\mathbf{q}}} \rho_q \left( \frac{1}{1 + \beta q^2} \right) \equiv - \frac{V_{\mathbf{q}}}{\hbar \omega_{\mathbf{q}}} \rho_q g_q,
\end{equation}
where $\beta$ is the variational parameter controlling the lattice recoil. To evaluate the Peierls--Yoccoz energy, we must calculate the non-diagonal matrix elements between a localized state centered at $\mathbf{R}/2$ and one centered at $-\mathbf{R}/2$. 

With the electronic degrees of freedom integrated out analytically, the phonon overlap exponent $G(\mathbf{R})$, the bare phonon potential energy $U_{\text{ph}}(\mathbf{R})$, and the electron--phonon interaction energy $V_{\text{int}}(\mathbf{R})$ are defined in 3D momentum space as:
\begin{align}
    G(\mathbf{R}) &= \sum_{\mathbf{q}} |f_{\mathbf{q}}|^2 \left( \mathrm{e}^{\mathrm{i}\mathbf{q}\cdot\mathbf{R}} - 1 \right), \\
    U_{\text{ph}}(\mathbf{R}) &= \sum_{\mathbf{q}} \hbar \omega_{\mathbf{q}} |f_{\mathbf{q}}|^2 \mathrm{e}^{\mathrm{i}\mathbf{q}\cdot\mathbf{R}}, \\
    V_{\text{int}}(\mathbf{R}) &= \sum_{\mathbf{q}} \left[ V_{\mathbf{q}} f_{\mathbf{q}}^* \mathrm{e}^{\mathrm{i}\mathbf{q}\cdot\mathbf{R}/2} \tilde{\rho}_{\mathbf{q}}(\mathbf{R}) + c.c. \right],
\end{align}
where $\tilde{\rho}_{\mathbf{q}}(\mathbf{R}) = \langle \phi_{\mathbf{R}/2} | \mathrm{e}^{\mathrm{i}\mathbf{q}\cdot\mathbf{r}} | \phi_{-\mathbf{R}/2} \rangle / \langle \phi_{\mathbf{R}/2} | \phi_{-\mathbf{R}/2} \rangle$ is the electronic transition density.

Because the system is isotropic, the angular integrals over the solid angle $d\Omega_q$ analytically map the complex exponential phases $\mathrm{e}^{\mathrm{i}\mathbf{q}\cdot\mathbf{R}}$ into spherical Bessel functions, $\int d\Omega_q \mathrm{e}^{\mathrm{i}\mathbf{q}\cdot\mathbf{R}} \propto \mathrm{sinc}(qR)$. Applying the continuum limit $\sum_{\mathbf{q}} \to \int d^3q / (2\pi)^3$, these expressions strictly reduce to stable 1D integrals over the scalar momentum $q$:
\begin{align}
    G(R) &= \frac{\alpha\sqrt{2}}{\pi}\int_0^\infty dq \, \rho_q^2 g_q^2 \left( \mathrm{sinc}(qR) - 1 \right), \\
    U_{\text{ph}}(R) &= \frac{\alpha\sqrt{2}}{\pi} \int_0^\infty dq \, \rho_q^2 g_q^2 \, \mathrm{sinc}(qR), \\
    V_{\text{int}}(R) &= -2\left(\frac{\alpha\sqrt{2}}{\pi}\right) \int_0^\infty dq \, (\rho_q g_q) \tilde{\rho}_{q}(R),
\end{align}
where $\tilde{\rho}_{q}(R)$ is the analytically accumulated, angle-averaged transition density that we define below. 

For the 3-Gaussian trial state, the un-normalized 3D transition density between basis functions $i$ and $j$ is given by:
\begin{equation}
    \tilde{\rho}_{ij}(\mathbf{q}, \mathbf{R}) = \int d^3r \, \phi_i(\mathbf{r}) \phi_j(\mathbf{r} - \mathbf{R}) \mathrm{e}^{\mathrm{i}\mathbf{q}\cdot\mathbf{r}}.
\end{equation}
Using the standard Gaussian product theorem, the product $\phi_i(\mathbf{r}) \phi_j(\mathbf{r} - \mathbf{R})$ forms a new Gaussian centered at $\mathbf{R}_c = (\mu_j / \mu_{ij})\mathbf{R}$, scaled by the real-space overlap $I_{ij}(R)$. Taking the Fourier transform yields:
\begin{equation}
    \tilde{\rho}_{ij}(\mathbf{q}, \mathbf{R}) = I_{ij}(R) \exp\left(-\frac{q^2}{2\mu_{ij}}\right) \exp\left(\mathrm{i}\mathbf{q}\cdot \frac{\mu_j}{\mu_{ij}}\mathbf{R}\right).
\end{equation}
Because the system is globally isotropic, we perform a spherical angular average over the relative orientation between $\mathbf{q}$ and $\mathbf{R}$. This maps the complex plane-wave phase directly into a spherical Bessel function. Summing over all Gaussian pairs and normalizing by the total real-space overlap gives the exact 1D radial transition density:
\begin{equation}
    \tilde{\rho}_q(R) = \frac{1}{\mathcal{N}} \sum_{i=1}^3 \sum_{j=1}^3 c_i c_j I_{ij}(R) \exp\left(-\frac{q^2}{2\mu_{ij}}\right) \mathrm{sinc}\left( \frac{\mu_i}{\mu_{ij}} q R \right).
\end{equation}
Here, the symmetry of the sum under the exchange of indices $i$ and $j$ allows the shift factor to be written as $\mu_i / \mu_{ij}$, mapping perfectly to the nested summation loops utilized in the numerical implementation. This exact analytical reduction ensures that the highly oscillatory cross-terms are evaluated accurately without requiring a dense 3D numerical grid.

\subsection{Scale-Invariant Quadrature}
The final PY energy requires integrating over the generator coordinate $R$:
\begin{equation}
    E_{\text{PY}} = \frac{\int dR \, R^2 \exp(G(R)) \left[ T_{\text{el}}(R) + I_{\text{el}}(R)U_{\text{ph}}(R) + V_{\text{int}}(R) \right]}{\int dR \, R^2 \exp(G(R)) I_{\text{el}}(R)}.
\end{equation}
A naive integration over $R$ and $q$ fails in the strong-coupling limit ($\alpha \gg 1$). As the polaron localizes, $\mu \to \infty$, causing the physical radius to shrink drastically. This causes the volume element $R^2 dR$ to approach numerical zero (e.g., $\sim 10^{-30}$), leading to catastrophic floating-point division errors.

To guarantee optimization stability, we introduce a scale-invariant quadrature grid. We define an effective width $\mu_{\text{eff}} = (\sum c_i^2 \mu_i) / (\sum c_i^2)$ and transform the integration variables to dimensionless coordinates:
\begin{equation}
    x = R \sqrt{\mu_{eff}}, \quad \quad y = q / \sqrt{\mu_{eff}}.
\end{equation}
Under this transformation, the Jacobian volume factor $\mu_{eff}^{-3/2}$ factors out of both the numerator and the denominator, analytically canceling the source of the numerical underflow.

The quadratures are evaluated on highly optimized, memory-efficient 500-point hybrid grids:
\begin{itemize}
    \item Scale-Invariant $x$-grid (Real Space): Concatenates a dense linear region ($x \in [0, 0.1]$, 150 points) to exactly resolve the fragile $R \to 0$ core where $\mathrm{sinc}(qR) \approx 1$, and a logarithmic tail ($x \in [0.101, 20.0]$, 350 points) to capture the exponentially decaying overlap without wasting array size.
    \item Scale-Invariant $y$-grid (Momentum Space): Concatenates a linear core ($y \in [0, 0.1]$, 100 points) to capture macroscopic $q \to 0$ polarization, and a deep logarithmic tail ($y \in [0.101, 3000.0]$, 400 points). The massive upper bound of $3000.0$ is required to adequately capture the broad momentum-space tails of the sharp, adiabatic Gaussian cores formed in the strong-coupling limit.
\end{itemize}

\section{Electron--phonon coupling matrix elements} \label{sec:elph_coupling_elements}
Our calculations require the computation of electron--phonon matrix elements $g_{mn}^{\nu}(\mathbf{k}, \mathbf{q})$. These matrix elements of the response of the self-consistent potential with respect to a perturbation of wave vector $u_{\mathbf{q},\kappa\alpha}$  are given in the phonon eigenmode basis by
\begin{align} \label{eq:elph_SI}
    g_{mn\nu}(\mathbf{k},\mathbf{q}) = \sqrt{\frac{\hbar}{2\omega_{\nu\mathbf{q}}}}\sum_{\kappa\alpha} \frac{e^{\nu}_{\kappa\alpha}(\mathbf{q})}{\sqrt{M_\kappa}} \langle \psi_{m\mathbf{k}+\mathbf{q}} | \partial_{\mathbf{q},\kappa\alpha}V_{\mathrm{SCF}} | \psi_{n\mathbf{k}}\rangle  
\end{align}
on arbitrarily fine $\mathbf{k}$- and $\mathbf{q}$-meshes. Here, $m,n$ label electronic Bloch states, $\nu$ labels phonon eigenmodes $e_{\kappa\alpha}^\nu(\mathbf{q})$, and $\kappa,\alpha$ label atoms in the unit cell and corresponding cartesian coordinates. 
These matrix elements are obtained by Wannier interpolation from an initial set of matrix elements computed on coarse $\mathbf{k}$- and $\mathbf{q}$-point meshes, denoted $\mathbf{k}_c$ and $\mathbf{q}_c$. After Fourier transforming the perturbation matrix elements to Wannier gauge, one obtains
\begin{align}
    g_{ij\kappa\alpha}(\mathbf{R}_\text{e}, \mathbf{R}_\text{p}) = \frac{1}{N_\text{e} N_\text{p}} \sum_{\mathbf{k}_\text{c} \mathbf{q}_\text{c} mn} \mathrm{e}^{-\mathrm{i} (\mathbf{k}_\text{c}\cdot \mathbf{R}_\text{e} + \mathbf{q}_\text{c} \cdot\mathbf{R}_\text{p})} \: \mathcal{U}_{im}^*(\mathbf{k}_\text{c}+\mathbf{q}_\text{c}) \: \langle \psi_{m\mathbf{k}_\text{c}+\mathbf{q}_\text{c}} | \partial_{q_c,\kappa\alpha}V_{\mathrm{SCF}} | \psi_{n\mathbf{k}_\text{c}}\rangle \: \mathcal{U}_{jn}(\mathbf{k}_\text{c}),
\end{align}
where $\tilde{i},\tilde{j}$ denote Wannier-gauge electronic indices, $i,j$ are coarse-grid Bloch-gauge indices. 
Wannier interpolation rests on the assumption that $g$ is sufficiently short-ranged in real space. If this is given, we can compute $g$ with arbitrary resolution in $k$ space via the inverse transformation
\begin{align}
    g_{mn\kappa\alpha}(\mathbf{k},\mathbf{q}) = \sum_{\mathbf{R}_\text{e} \mathbf{R}_\text{p} ij} \mathrm{e}^{-\mathrm{i} (\mathbf{k} \cdot \mathbf{R}_\text{e} + \mathbf{q} \cdot\mathbf{R}_\text{p})} U_{im}^*(\mathbf{k}+\mathbf{q}) U_{jn}(\mathbf{k})\: g_{ij\kappa\alpha}(\mathbf{R}_\text{e}, \mathbf{R}_\text{p}).
\end{align}
Following the notation in the main text, $U_{jn}(\mathbf{k})$ are obtained by diagonalizing the interpolated electronic Hamiltonian in Wannier gauge. 

In polar materials, electron--phonon matrix elements generally are not short-ranged, as strong lattice dressing induces a macroscopic electric field. In these cases, one is forced to correct the missing long-range contribution analytically\cite{vogl_microscopic_1976,verdi_frohlich_2015}. Corrections to the electron--phonon matrix elements are derived from a multipole expansion which truncates after the quadrupole term. The dipole correction is evaluated as 
\begin{align} \label{eq:longrange}
    g_{mn\kappa\alpha}^{L} (\mathbf{k},\mathbf{q}) = \sum_{\mathbf{G}\neq-\mathbf{q}} \frac{[(\mathbf{q}+\mathbf{G}) \cdot Z_\kappa^*]_\alpha \mathrm{e}^{-\mathrm{i}\boldsymbol{\tau}_\kappa \cdot (\mathbf{q}+\mathbf{G})}}{(\mathbf{q}+\mathbf{G}) \cdot \epsilon \cdot (\mathbf{q}+\mathbf{G})} \langle \psi_{m\mathbf{k}+\mathbf{q}}| \mathrm{e}^{\mathrm{i}(q+\mathbf{G}) \cdot \mathbf{r}}|\psi_{n\mathbf{k}}\rangle,
\end{align}
where $\boldsymbol{\tau}_\kappa$ is the position of atom $\kappa$ in the unit cell and $Z_\kappa^*$ are Born effective charges. The sum over reciprocal lattice vectors $\mathbf{G}$ in \cref{eq:longrange} may be performed via the Ewald method, truncating the sum by introducing an exponential dampening factor. This is physically motivated by keeping the long-range contribution actually long-ranged, seeing as it should be dominated by contributions from small $\mathbf{q}$ vectors. The same reasoning allows us to approximate~\cite{sjakste_wannier_2015}
\begin{align}
    \langle \psi_{m\mathbf{k}+\mathbf{q}}| \mathrm{e}^{\mathrm{i}(\mathbf{q}+\mathbf{G}) \cdot \mathbf{r}}|\psi_{m\mathbf{k}}\rangle = \langle u_{m\mathbf{k}+\mathbf{q}}| \mathrm{e}^{\mathrm{i}\mathbf{G} \cdot \mathbf{r}}|u_{n\mathbf{k}}\rangle  =\sum_{ij} U_{im}(\mathbf{k}+\mathbf{q}) U_{jn}^*(\mathbf{k}) \underbrace{\langle u_{i\mathbf{k}+\mathbf{q}}| \mathrm{e}^{\mathrm{i}\mathbf{G} \cdot \mathbf{r}}|u_{j\mathbf{k}}\rangle}_{\approx \delta_{ij}}.
\end{align}
Here, we used Bloch's theorem $|\psi_{m\mathbf{k}}\rangle = \exp(\mathrm{i} \mathbf{k} \cdot \mathbf{r}) |u_{m\mathbf{k}}\rangle$.
The final electron--phonon coupling matrix elements are then given by 
\begin{equation}
    g_{ij\kappa\alpha}(\mathbf{k},\mathbf{q}) = g_{ij\kappa\alpha}^S(\mathbf{k},\mathbf{q}) + g^L_{ij\kappa\alpha}(\mathbf{k},\mathbf{q}),
\end{equation}
where the short-ranged contribution $g_{ij\kappa\alpha}^S(\mathbf{k},\mathbf{q})$ is obtained via Wannier-interpolation of the electron--phonon matrix elements obtained from DFPT on the coarse $\mathbf{k}$-mesh.
For our present approach, we finally transform $g_{ij\kappa\alpha}(\mathbf{k},\mathbf{q})$ into phonon eigenmode basis. 

\subsection{Alignment of Degenerate Modes}
We found that in order to converge our variational calculations efficiently and produce a reliable finite-size extrapolation degenerate modes need to be aligned between $\mathbf{q}$ and $-\mathbf{q}$ when generating the electron--phonon coupling elements. 

For a non-magnetic crystal, dynamical matrices (and therefore phonon eigenmodes) obey $D(\mathbf{q}) = D^*(-\mathbf{q})$. Diagonalizing these dynamical matrices breaks this relation, as non-degenerate eigenvectors can pick up an arbitrary $U(1)$ phase, while each $n$-fold degenerate subspace is determined only up to an arbitrary $U(n)$ rotation. To restore the relationship between phonon modes at $\mathbf{q}$ and $\mathbf{-q}$ we perform an orthogonal Procrustes alignment for each pair of $\mathbf{q}$ and $\mathbf{-q}$. Following the formal solution of the orthogonal Procrustes problem, we define the overlap
\begin{align}
    O_{\mu \nu} = \sum_{\kappa\alpha} e_{\kappa\alpha}^{\mu}(-\mathbf{q})\: M_{\kappa} \: e_{\kappa\alpha}^{\nu}(\mathbf{q}),
\end{align}
where $\mu,\nu$ index into a block of degenerate phonon modes.
We are interested to find the rotation $W$, such that
\begin{equation}
    W = \arg\min_{\tilde{W}} ||\tilde{W}_{\mu\nu} \: \left(e^{\nu}_{\kappa\alpha}(-\mathbf{q})\right)^* - e^{\mu}_{\kappa\alpha}(\mathbf{q})||_F^2 
\end{equation}
is minimized over all unitary $\tilde{W}$'s. Here, $||\cdot||_F^2$ is the Frobenius norm. This is formally solved by performing an SVD for $O = U \Sigma V^\dagger$, leading to 
\begin{equation}
    W = U V^\dagger.
\end{equation}
For each pair $(\mathbf{q}, -\mathbf{q})$ we then rotate according to
\begin{equation}
    e_{\kappa\alpha}^{\nu}(-\mathbf{q}) \longleftarrow\sum_{\nu} e_{\kappa\alpha}^{\mu}(-\mathbf{q}) W_{\mu \nu}^*.
\end{equation}
For a nondegenerate mode the $O$-block reduces to a scalar and the rotation collapses to a simple phase alignment,
\begin{align}
    e_{\kappa\alpha}^{\nu}(-\mathbf{q}) \longleftarrow  e_{\kappa\alpha}^{\nu}(-\mathbf{q}) \frac{O^*}{|O|}.
\end{align}
We further symmetrize modes and phonon frequencies via
\begin{align}
    e_{\kappa\alpha}^{\nu}(\mathbf{q}) &\longleftarrow \frac{1}{2}\big( e_{\kappa\alpha}^{\nu}(\mathbf{q})\:+\: [e_{\kappa\alpha}^{\nu}(-\mathbf{q})]^* \big),\\
    \omega_{\nu}(\mathbf{q}) &\longleftarrow \frac{1}{2}\big( \omega_{\nu}(\mathbf{q})\:+\: \omega_{\nu}(-\mathbf{q}) \big)
\end{align}
These aligned modes and frequencies then enter \cref{eq:elph_SI}.

\section{Efficient contraction path}\label{sec:contract}
Substituting Eq.~\eqref{eq:g_svd} into the last term of Eq.~\eqref{eq:dd2_var_energy}, 
we obtain
\begin{align}
    \begin{split}
        &\frac{2}{\sqrt{N_k}} \operatorname{Re} \Bigg\{
        \sum_{i\mathbf{k}} D_{\mathbf{k}} \tilde{A}_{i\mathbf{k}}
        \sum_{\nu\mathbf{q}} L_{\nu}(\mathbf{q}) B_{\nu,-\mathbf{q}}^* \tilde{A}_{i,\mathbf{k}+\mathbf{q}} \\
        & + \sum_{ij\nu\gamma\mathbf{q}} V_{ij\nu}^\gamma(\mathbf{q}) B_{\nu,-\mathbf{q}}^*
        \sum_\mathbf{k}\left[\Sigma_{ij}^\gamma(\mathbf{k}) D_{\mathbf{k}} \tilde{A}_{j\mathbf{k}}\right]
        \tilde{A}_{i,\mathbf{k}+\mathbf{q}}^*
        \Bigg\},
    \end{split}
\end{align}
where we introduced the variational electronic amplitudes in the Wannier basis,
\begin{align}
    \tilde{A}_{i\mathbf{k}} = \sum_{n} U_{in}(\mathbf{k}) A_{n\mathbf{k}}.
\end{align}
The innermost contraction over $\mathbf {k} $ can be performed with a fast Fourier transform as explained in the main text, achieving significant cost reduction in $N_{\mathbf k}$.

\section{SVD error} \label{sec:svd_error}
\begin{figure}
    \centering
    \includegraphics[width=\linewidth]{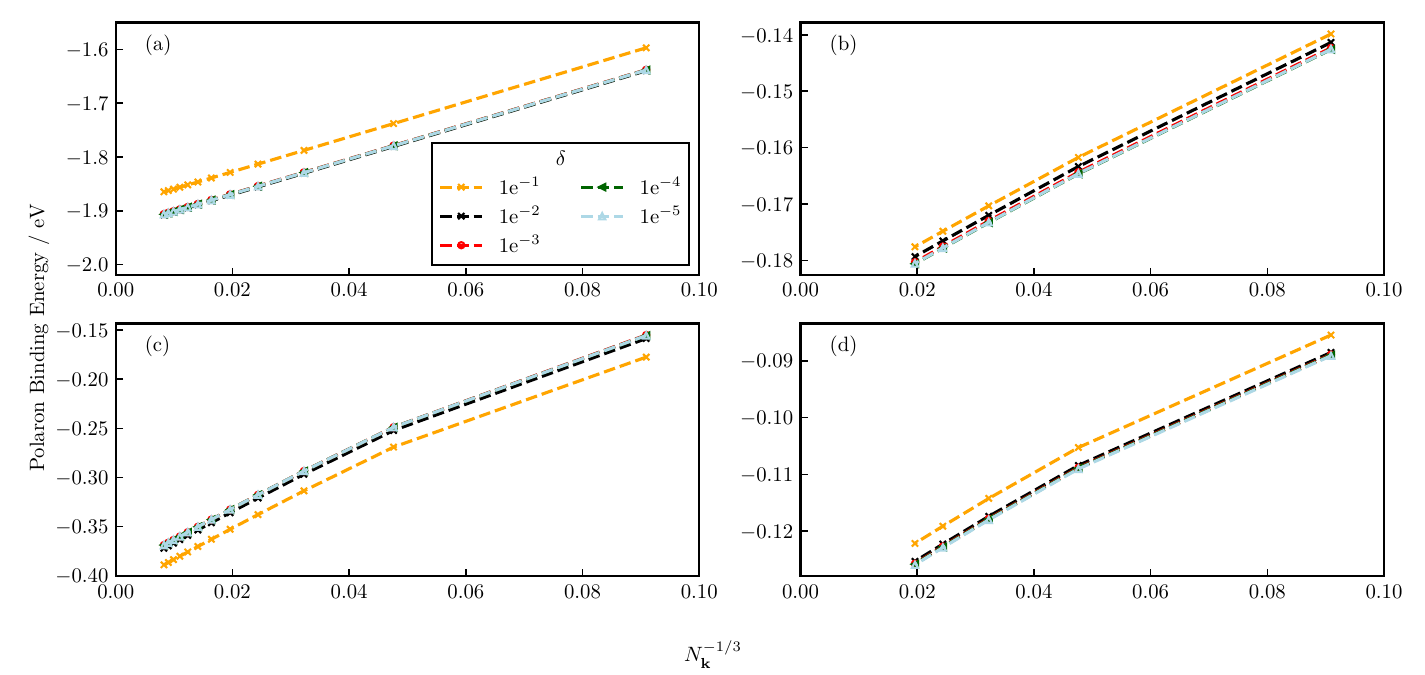}
    \caption{Size extrapolation for various SVD truncation thresholds $\delta$, computed for (a) electron and (b) hole polarons in \ce{LiF}, and electron polarons in (c) anatase and (d) rutile.}
    \label{fig:svd_binding}
\end{figure}

For a systematic analysis of the error introduced by truncating the singular value decomposition introduced in \cref{eq:g_svd}, we scan over five different truncation thresholds $\delta \in \{1\mathrm{e}^{-1}, 1\mathrm{e}^{-2}, 1\mathrm{e}^{-3}, 1\mathrm{e}^{-4}, 1\mathrm{e}^{-5}\}$, and compute the corresponding polaron binding energies. The extrapolated values and convergence plots are shown in Tab.~\ref{tab:formation_energies} and Fig.~\ref{fig:svd_binding}, respectively. 

\begin{table}[h]
    \centering
    {\setlength{\tabcolsep}{8pt}
    \begin{tabular}{lccccc}
     \hline
     \hline
      SVD threshold & $1\mathrm{e}^{-1}$& $1\mathrm{e}^{-2}$ & $1\mathrm{e}^{-3}$ & $1\mathrm{e}^{-4}$ & $1\mathrm{e}^{-5}$ \\ [0.8ex]
     \hline
     LiF (e) & -0.41535 & --0.39844 & --0.39521  & --0.39524 & --0.39524  \\
     LiF (h) &  1.89018 & 1.93442 & 1.93275 & 1.93323  &  1.93329 \\
     Anatase & --0.13459 & --0.13774 & --0.13815 &  --0.13824 & --0.13835 \\
     Rutile  & --0.18892 & --0.19073 & --0.19171 &--0.19201 & --0.19196 \\ [0.5ex]
     \hline
     \hline
    \end{tabular}}
    \caption{Extrapolated dD2 polaron formation energies in \si{\electronvolt} for different SVD thresholds.}
    \label{tab:formation_energies}
\end{table}

\begin{figure}
    \centering
    \includegraphics[width=\linewidth]{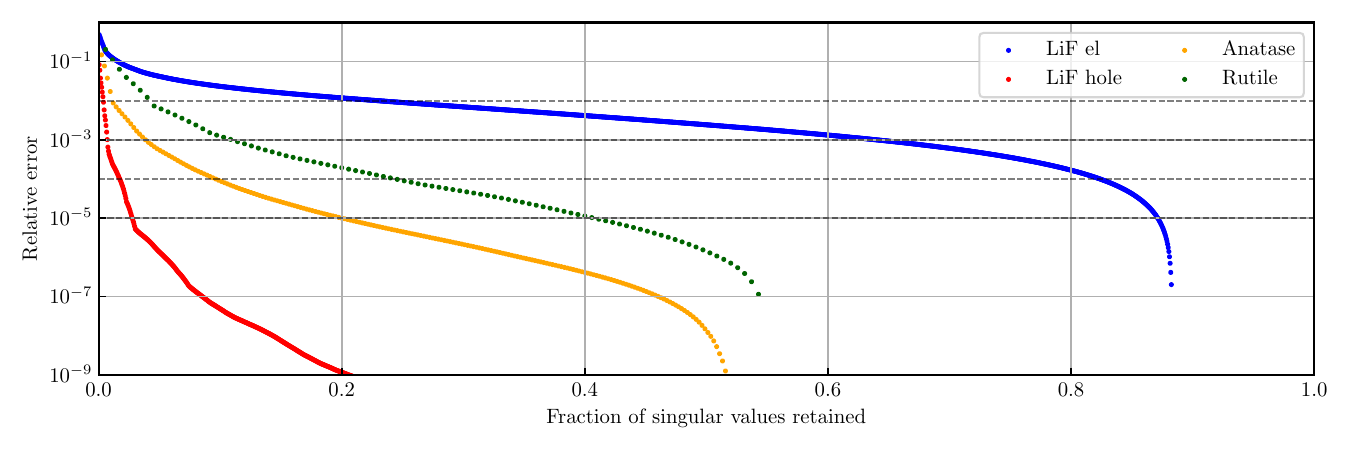}
    \caption{Relative error in the reconstruction of $g_{mn\nu}(\mathbf{k},\mathbf{q})$ from the corresponding number of largest singular values retained in the tensor decomposition.}
    \label{fig:svd_err}
\end{figure}

Finally, Fig.~\ref{fig:svd_err} shows the relative error associated with a truncation of the corresponding number of singular values. This relative error is defined as 
\begin{align}
     \frac{\sum_{ij} \sum_{\gamma>N_c} \sigma^2_{ij\gamma}}{\sum_{ij}\sum_\gamma \sigma^2_{ij\gamma}} = 1- \frac{\sum_{ij}\sum_{\gamma\leq N_c} \sigma^2_{ij\gamma}}{\sum_{ij}\sum_\gamma \sigma^2_{ij\gamma}} = \frac{\sum_{\mathbf{k}\mathbf{q}mn\nu}|g_{mn\nu}(\mathbf{k},\mathbf{q}) - g^{N_c}_{mn\nu}(\mathbf{k},\mathbf{q})|^2}{\sum_{\mathbf{k}\mathbf{q}mn\nu}|g_{mn\nu}(\mathbf{k},\mathbf{q}) |^2},
\end{align}
where $g^{N_c}_{mn\nu}(\mathbf{k},\mathbf{q}) = \sum_{ij}\sum_{\gamma\leq N_c} U_{mi}^*(\mathbf{k}+\mathbf{q}) \sigma_{ij\gamma} \tilde{\Sigma}_{ij}^\gamma(\mathbf{k}) V_{ij\nu}^\gamma(\mathbf{q}) U_{nj}(\mathbf{k})$, with singular values $ \sigma_{ij\gamma}$ and singular vectors $\tilde{\Sigma}_{ij}^\gamma(\mathbf{k})$ and $ V_{ij\nu}^\gamma(\mathbf{q})$. In the main text we absorbed the singular values in the singular vectors such that $\sigma_{ij\gamma}\tilde{\Sigma}_{ij}^\gamma(\mathbf{k}) = \Sigma_{ij}^\gamma(\mathbf{k})$.

The number of singular values retained to obtain the results presented in the main text, corresponding to a relative error of $1\mathrm{e}^{-3}$, are 1255 for the \ce{LiF} electron polaron, 15 for the \ce{LiF} hole polaron, 17 for the anatase electron polaron, and 20 for the rutile electron polaron. 

\section{Size extrapolations} \label{sec:SI_size_extra}
\begin{figure}
    \centering
    \includegraphics[width=\linewidth]{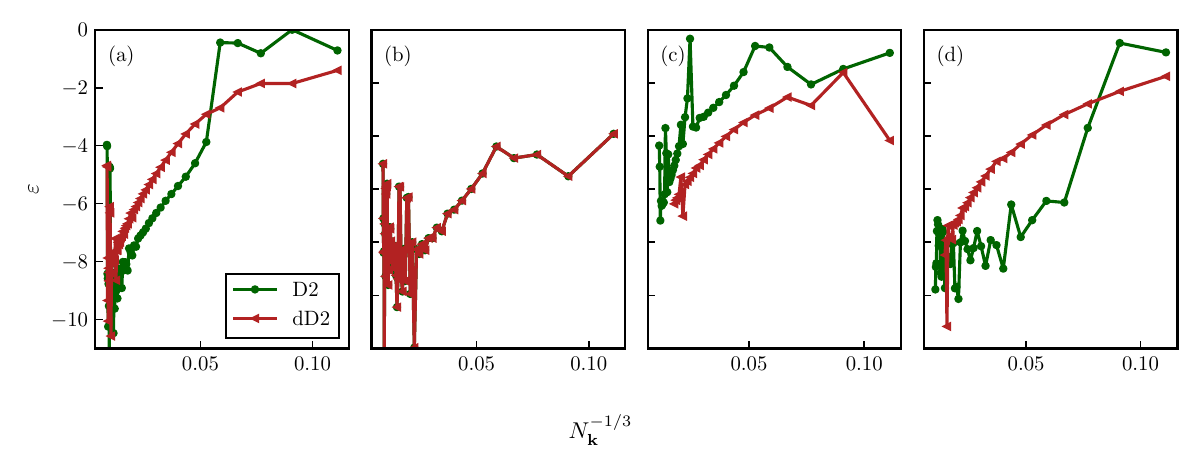}
    \caption{Error due to unexplained variance for (a) the electron and (b) hole polaron in \ce{LiF}, and the electron polarons in (c) anatase and (d) rutile.}
    \label{fig:unexpl_var}
\end{figure}

Unexplained variance may be used to verify the quality of convergence. Here, {\it unexplained} is a technical term for the error not explained by the linear fit. We are however able to attribute this error to a finite convergence threshold. For visualization purposes we compute its logarithm,
\begin{align}
    \varepsilon& = \log_{10}(1 - R^2) \\
    &= \log_{10}\left(\frac{\sum_{N_{\mathbf{k}}} (E_{N_{\mathbf{k}}} - y_{N_{\mathbf{k}}})^2}{\sum_{{N_{\mathbf{k}}}} (E_{N_{\mathbf{k}}} - \bar{E}_{N_{\mathbf{k}}})^2}\right),
\end{align}
where $\bar{E}_{N_{\mathbf{k}}} = N^{-1} \sum_{{N_{\mathbf{k}}}} E_{N_{\mathbf{k}}}$ and $N$ is the number of samples. 
For data following a perfectly linear function $y_{N_{\mathbf{k}}}$, $\varepsilon$ will go towards $-\infty$. We compute $\varepsilon$ from three neighboring $N_\mathbf{k}$ values, starting from results for $\mathbf{k}$-meshes $3^3$ to $7^3$, going up to $117^3$ to $121^3$ for \ce{LiF}, $65^3$ to $69^3$ for rutile, and $53^3$ to $57^3$ for anatase. Results are shown in \cref{fig:unexpl_var}. The $y$-axis corresponds to the maximum value of $N_\mathbf{k}^{-1/3}$ used in the 3 point sampling.

\begin{figure}
    \centering
    \includegraphics[width=\linewidth]{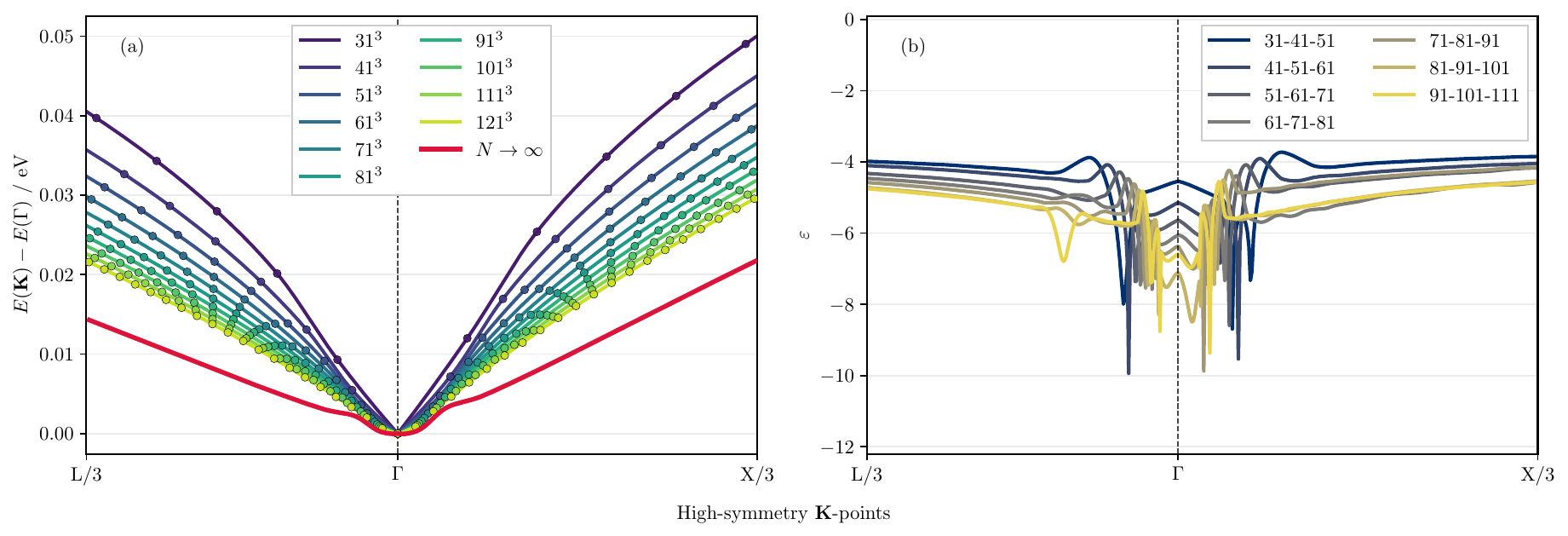}
    \caption{(a) Finite size extrapolation for the \ce{LiF} electron polaron band. Solid lines correspond to the PCHIP interpolant, dots correspond to the dD2 ground truth. (b) Error due to unexplained variance for linear fits through the ground state on 3 specified $\mathbf{k}$-meshes. GDM convergence threshold is chosen to be $1\mathrm{e}^{-3}$.}
    \label{fig:band_error_analysis}
\end{figure}

We conduct a similar error analysis for the size extrapolation of the polaron band displayed in \cref{fig:bandstructure}. In order to obtain the \ce{LiF} electron polaron band in the thermodynamic limit we first use a piecewise cubic Hermite interpolating polynomial (PCHIP) interpolation to resolve the polaron band across the high-symmetry path displayed in \cref{fig:band_error_analysis} at arbitrary $\mathbf{K}$-points. We then sample for each band obtained on a certain $\mathbf{k}$-mesh the same set of $\mathbf{K}$-points in order to do a linear size extrapolation as done previously for each of these sampled $\mathbf{K}$-points.  

Finally, to propagate the DiagMC errors to the thermodynamic limit as reported in \cref{tab:formation_energies_P}, we fit the binding energies $E(N_{\mathbf{k}_i})$ at the largest three supercells to the same linear form used for all other size extrapolations in this work. We did so via weighted least squares with weights $w_i = 1 / \sigma_i^2$, where $\sigma_i$ is the DiagMC error at $N_{\mathbf{k}_i}$. The corresponding model can be written as $y = X\beta + \varepsilon$, where $X_i = (1, 1/N_{\mathbf{k}_i}^{-1/3})$ and $\beta=(E_\infty,a)^T$. The covariance of the best-fit parameters, obtained from propagating the input $\sigma_i$ through this linear model, is given by $(X^T W X)^{-1}$ with $W=\mathrm{diag}(w_i)$. We obtain the extrapolated DiagMC error bars by taking the square root of the first diagonal entry of the covariance.

\section{More details on the comparison of DiagMC and dD2 polaron bands}\label{sec:bandbias}
\begin{table}[h]
    \centering
    {\setlength{\tabcolsep}{8pt}
    \begin{tabular}{lccccc}
     \hline
     \hline
     & dD2 1-band & DiagMC & dD2 4-band & dD2 1-band & dD2 1-band \\ [0.5ex]
     \hline
     $N_\text{coarse}$ &$12^3$& $6^3$ & $6^3$ & $6^3$ & $6^3$ \\
     $N_c$ & 1255 & 20 & 20 & 101 & 20 \\ 
     $\delta$ & $1\mathrm{e}^{-3}$ & $7.1\mathrm{e}^{-3}$ & $7.1\mathrm{e}^{-3}$ & $1\mathrm{e}^{-3}$ & $7.1\mathrm{e}^{-3}$\\
     \hline
     $E(\mathbf{K})$ / eV  & --0.35051 &--0.34525 & --0.33953 & --0.33508 & --0.32947\\
     [0.5ex]
     \hline
     \hline
    \end{tabular}}
    \caption{Polaron binding energies for the \ce{LiF} electron polaron on a $120^3$ $\mathbf{k}$-mesh, at crystal momentum $\mathbf{K} = (\frac{16}{120}, \frac{16}{120}, \frac{16}{120})$ obtained from matrix elements generated from initial DFT and DFPT coarse meshes of size $N_\text{coarse} = 6^3$ and $N_\text{coarse} = 12^3$. $\delta$ is the relative SVD truncation threshold as defined in \cref{sec:svd_error}.}
    \label{tab:diagmc_band_diff}
\end{table}

In the main text, we showed that the dD2 polaron band on a $120^3$ $\mathbf{k}$-mesh finds a lower energy than the reported DiagMC polaron band away from $\Gamma$. 
While determining the root cause of the seemingly biased DiagMC polaron band structure is beyond the scope of our work, we suspect that it mainly originates from errors in approximating the eph Hamiltonian.
For our study, we have performed our own density functional perturbation theory (DFPT) calculations over a coarse grid of $N_\text{coarse}=12^3$, whereas Luo \emph{et al.} used $N_\text{coarse}=6^3$~\cite{luo_first-principles_2025}.
Another difference is that our criterion for determining the number of singular vectors retained is more stringent than theirs, leading to a much larger number of singular vectors and, in the end, a smaller compression error.
With these two factors combined, our Hamiltonian matrix elements are numerically more accurate.

To gauge the associated band energy error with Luo \emph{et al.}'s Hamiltonian, we evaluated dD2 energies at $\mathbf{K} = (\frac{16}{120}, \frac{16}{120}, \frac{16}{120})$, where the DiagMC errorbars are comparably small. We additionally performed dD2 using the Hamiltonian matrix elements provided by Luo \emph{et al.}, using $N_{c} = 20$, and (i) a projection onto the lowest Bloch band, and (ii) the optimized dD2 energy for the inclusion of all 4 Bloch bands in the optimization. 
These results are presented in \cref{tab:diagmc_band_diff} along with the corresponding dD2 band energy reported in the main text (using 12$^3$ and $N_c = 1255$) and DiagMC energy.
As shown in \cref{tab:diagmc_band_diff}, the dD2 energy difference coming from the inaccurate eph Hamiltonian matrix element is about 0.01 eV (4-band)  or 0.02 eV (1-band), which is larger than the difference ($\simeq$ 5 meV) between dD2 and DiagMC observed in the polaron band.
Furthermore, the subsequent dD2 energy with the same Hamiltonian matrix elements as DiagMC is higher than DiagMC, indicating that DiagMC energy is better (in the variational sense) than dD2.

We can decompose the 1-band Hamiltonian error into two contributions. Error introduced by the coarse DFT/DFPT $\mathbf{k}$-mesh sampling amounts to $\sim15$ meV. This is seen by comparing dD2 1-band $12^3$ and $6^3$ with $\delta = 1e^{-3}$.
Additionally, the dD2 1-band energy difference between different SVD thresholds predicts that the SVD truncation error is $\sim6$ meV.
Overall, the dD2 1-band energy difference due to the Hamiltonian error is $\sim 21$ meV.
As the difference between the DiagMC results and dD2 1-band results with $\delta=1\mathrm{e}^{-3}$ is only $\sim5$meV, it is quite likely that the accuracy of Hamiltonian error is playing a more important role than the sampling error of DiagMC (if any) in \cref{fig:bandstructure}.

\section{dD2 gradient and hessian} \label{sec:derivatives}
\subsection{Gradient}
We derive the analytical gradients for the dD2 wavefunction below. The corresponding D2 expressions are obtained immediately by setting \(D_{\mathbf{k}}=1\). These derivatives assume a Hermitian electron--phonon coupling tensor. If the full tensor \(g_{mn\nu}(\mathbf{k},\mathbf{q})\) is stored explicitly, hermiticity can be enforced directly. Otherwise, for example when working with singular vectors obtained from a non-Hermitian real-space tensor \(g_{ij\kappa\alpha}(\mathbf{R}_{\mathrm e},\mathbf{R}_{\mathrm p})\), one may hermitianize the coupling on the fly by defining
\begin{equation}
    g_{mn\nu}(\mathbf{k},\mathbf{q})
    = \frac{1}{2}\Bigl(
        \tilde{g}_{mn\nu}(\mathbf{k},\mathbf{q})
        + \tilde{g}^{*}_{nm\nu}(\mathbf{k}+\mathbf{q},-\mathbf{q})
      \Bigr),
\end{equation}
and then inserting the corresponding low-rank decompositions for
\(\tilde{g}_{mn\nu}(\mathbf{k},\mathbf{q})\).

Since both the electronic and bosonic variational parameters are generally complex in the dD2 wavefunction, one must evaluate four sets of derivatives:
\begin{subequations}
\begin{align}
\frac{\partial E_{\mathrm{dD2}}}{\partial \operatorname{Re}\{A_{n\mathbf{k}}\}}
&= 2 \langle \Psi | \Psi \rangle^{-1}
\Biggl[
    \operatorname{Re}\{A_{n\mathbf{k}}\}
    \Bigl(\varepsilon_{n\mathbf{k}} - E_{\mathrm{dD2}}(A,B)\Bigr)
    D_{\mathbf{k}}
    + \operatorname{Re}\{A_{n\mathbf{k}}\}
    \sum_{\nu\mathbf{q}} \omega_{\nu\mathbf{q}} |B_{\nu\mathbf{q}}|^2
    D_{\mathbf{k}+\mathbf{q}}
\nonumber\\
&\qquad\qquad
    + \operatorname{Re}\Biggl\{
        \sum_{m\nu\mathbf{q}}
        g_{nm\nu}(\mathbf{k}-\mathbf{q},\mathbf{q})
        A_{m\mathbf{k}-\mathbf{q}}
        B_{\nu,-\mathbf{q}}^{*}
        D_{\mathbf{k}-\mathbf{q}}
        +
        \sum_{m\nu\mathbf{q}}
        g_{mn\nu}(\mathbf{k},\mathbf{q})
        A_{m\mathbf{k}+\mathbf{q}}^{*}
        B_{\nu,-\mathbf{q}}^{*}
        D_{\mathbf{k}}
    \Biggr\}
\Biggr],
\\[0.5em]
\frac{\partial E_{\mathrm{dD2}}}{\partial \operatorname{Im}\{A_{n\mathbf{k}}\}}
&= 2 \langle \Psi | \Psi \rangle^{-1}
\Biggl[
    \operatorname{Im}\{A_{n\mathbf{k}}\}
    \Bigl(\varepsilon_{n\mathbf{k}} - E_{\mathrm{dD2}}(A,B)\Bigr)
    D_{\mathbf{k}}
    + \operatorname{Im}\{A_{n\mathbf{k}}\}
    \sum_{\nu\mathbf{q}} \omega_{\nu\mathbf{q}} |B_{\nu\mathbf{q}}|^2
    D_{\mathbf{k}+\mathbf{q}}
\nonumber\\
&\qquad\qquad
    - \operatorname{Re}\Biggl\{
        \mathrm{i}
        \sum_{m\nu\mathbf{q}}
        g_{nm\nu}(\mathbf{k}-\mathbf{q},\mathbf{q})
        A_{m\mathbf{k}-\mathbf{q}}
        B_{\nu-\mathbf{q}}^{*}
        D_{\mathbf{k}-\mathbf{q}}
        -
        \mathrm{i}
        \sum_{m\nu\mathbf{q}}
        g_{mn\nu}(\mathbf{k},\mathbf{q})
        A_{m\mathbf{k}+\mathbf{q}}^{*}
        B_{\nu-\mathbf{q}}^{*}
        D_{\mathbf{k}}
    \Biggr\}
\Biggr],
\\[0.5em]
\frac{\partial E_{\mathrm{dD2}}}{\partial \operatorname{Re}\{B_{\nu\mathbf{q}}\}}
&= 2 \langle \Psi | \Psi \rangle^{-1}
\Biggl[
    \operatorname{Re}\{B_{\nu\mathbf{q}}\}
    \sum_{n\mathbf{k}}
    \Bigl(
        \varepsilon_{n\mathbf{k}} + \omega_{\nu\mathbf{q}} - E_{\mathrm{dD2}}(A,B)
    \Bigr)
    |A_{n\mathbf{k}}|^2
    D_{\mathbf{k}+\mathbf{q}}
\nonumber\\
&\qquad\qquad
    + \operatorname{Re}\{B_{\nu\mathbf{q}}\}
    \sum_{n\mathbf{k}} |A_{n\mathbf{k}}|^2
    \sum_{\nu'\mathbf{q}'}
    \omega_{\nu'\mathbf{q}'}
    |B_{\nu'\mathbf{q}'}|^2
    D_{\mathbf{q}'+\mathbf{k}+\mathbf{q}}
\nonumber\\
&\qquad\qquad
    + \operatorname{Re}\Biggl\{
        \sum_{nm\mathbf{k}}
        g_{nm\nu}(\mathbf{k},-\mathbf{q})
        A_{n\mathbf{k}-\mathbf{q}}^{*}
        A_{m\mathbf{k}}
        D_{\mathbf{k}}
        +
        2 \operatorname{Re}\{B_{\nu\mathbf{q}}\}
        \sum_{n\mathbf{k}}
        D_{\mathbf{q}+\mathbf{k}}
        A_{n\mathbf{k}}
        \Biggl(
            \sum_{m\nu'\mathbf{q}'}
            g_{mn\nu'}(\mathbf{k},\mathbf{q}')
            A_{m\mathbf{k}+\mathbf{q}'}^*
            B_{\nu'-\mathbf{q}'}^{*}
        \Biggr)
    \Biggr\}
\Biggr],
\\[0.5em]
\frac{\partial E_{\mathrm{dD2}}}{\partial \operatorname{Im}\{B_{\nu\mathbf{q}}\}}
&= 2 \langle \Psi | \Psi \rangle^{-1}
\Biggl[
    \operatorname{Im}\{B_{\nu\mathbf{q}}\}
    \sum_{n\mathbf{k}}
    \Bigl(
        \varepsilon_{n\mathbf{k}} + \omega_{\nu\mathbf{q}} - E_{\mathrm{dD2}}(A,B)
    \Bigr)
    |A_{n\mathbf{k}}|^2
    D_{\mathbf{k}+\mathbf{q}}
\nonumber\\
&\qquad\qquad
    + \operatorname{Im}\{B_{\nu\mathbf{q}}\}
    \sum_{n\mathbf{k}} |A_{n\mathbf{k}}|^2
    \sum_{\nu'\mathbf{q}'}
    \omega_{\nu'\mathbf{q}'}
    |B_{\nu'\mathbf{q}'}|^2
    D_{\mathbf{q}'+\mathbf{k}+\mathbf{q}}
\nonumber\\
&\qquad\qquad
    + \operatorname{Re}\Biggl\{
        \sum_{nm\mathbf{k}}
        g_{nm\nu}(\mathbf{k},-\mathbf{q})
        A_{n\mathbf{k}-\mathbf{q}}^{*}
        A_{m\mathbf{k}}
        D_{\mathbf{k}}
        +
        2 \operatorname{Im}\{B_{\nu\mathbf{q}}\}
        \sum_{n\mathbf{k}}
        D_{\mathbf{q}+\mathbf{k}}
        A_{n\mathbf{k}}
        \Biggl(
            \sum_{m\nu'\mathbf{q}'}
            g_{mn\nu'}(\mathbf{k},\mathbf{q}')
            A_{m\mathbf{k}+\mathbf{q}'}^{*}
            B_{\nu'-\mathbf{q}'}^{*}
        \Biggr)
    \Biggr\}
\Biggr].
\end{align}
\end{subequations}

Upon inserting the decomposition of \(g_{mn\nu}(\mathbf{k},\mathbf{q})\) from \cref{eq:g_svd}, the electron--phonon contributions to the gradients can be written as
\begin{subequations}\label{eq:grad_ep_terms}
\begin{equation}
\begin{split}
\sum_{m\nu\mathbf{q}}
g_{nm\nu}(\mathbf{k}-\mathbf{q},\mathbf{q})
A_{m\mathbf{k}-\mathbf{q}}
B_{\nu-\mathbf{q}}^*
D_{\mathbf{k}-\mathbf{q}}
={}&
\sum_{\tilde{i}\tilde{j}}
U_{\tilde{i}n}^*(\mathbf{k})
\sum_{\gamma}\sum_{\mathbf{q}}
D_{\mathbf{k}-\mathbf{q}}
\Sigma_{\tilde{i}\tilde{j}}^\gamma(\mathbf{k}-\mathbf{q})
\\
&\times
\Biggl(
\sum_{\nu}
V_{\tilde{i}\tilde{j}\nu\gamma}(\mathbf{q})
B_{\nu-\mathbf{q}}^*
\Biggr)
\Biggl(
\sum_m
U_{\tilde{j}m}(\mathbf{k}-\mathbf{q})
A_{m\mathbf{k}-\mathbf{q}}
\Biggr)
\\
&+
\sum_{\tilde{i}}
U_{\tilde{i}n}^*(\mathbf{k})
\sum_{\mathbf{q}}
D_{\mathbf{k}-\mathbf{q}}
\Biggl(
\sum_m
U_{\tilde{i}m}(\mathbf{k}-\mathbf{q})
A_{m\mathbf{k}-\mathbf{q}}
\Biggr)
\Biggl(
\sum_{\nu}
L_{\nu}(\mathbf{q})
B_{\nu-\mathbf{q}}^*
\Biggr).
\end{split}
\end{equation}

\begin{equation}
\begin{split}
D_{\mathbf{k}}
\sum_{m\nu\mathbf{q}}
g_{mn\nu}(\mathbf{k},\mathbf{q})
A_{m\mathbf{k}+\mathbf{q}}^*
B_{\nu-\mathbf{q}}^*
={}&
D_{\mathbf{k}}
\sum_{\tilde{i}\tilde{j}\gamma}
\Sigma_{\tilde{i}\tilde{j}}^\gamma(\mathbf{k})
\,U_{\tilde{j}n}(\mathbf{k})
\sum_{\mathbf{q}}
\Biggl(
\sum_m
U_{\tilde{i}m}^*(\mathbf{k}+\mathbf{q})
A_{m\mathbf{k}+\mathbf{q}}^*
\Biggr)
\\
&\times
\Biggl(
\sum_{\nu}
B_{\nu-\mathbf{q}}^*
V_{\tilde{i}\tilde{j}\nu\gamma}(\mathbf{q})
\Biggr)
\\
&+
D_{\mathbf{k}}
\sum_{\tilde{i}}
U_{\tilde{i}n}(\mathbf{k})
\sum_{\mathbf{q}}
\Biggl(
\sum_m
U_{\tilde{i}m}^*(\mathbf{k}+\mathbf{q})
A_{m\mathbf{k}+\mathbf{q}}^*
\Biggr)
\Biggl(
\sum_{\nu}
L_{\nu}(\mathbf{q})
B_{\nu-\mathbf{q}}^*
\Biggr).
\end{split}
\end{equation}

\begin{equation}
\begin{split}
\sum_{nm\mathbf{k}}
g_{nm\nu}(\mathbf{k},-\mathbf{q})
A_{n\mathbf{k}-\mathbf{q}}^*
A_{m\mathbf{k}}
D_{\mathbf{k}}
={}&
\sum_{\tilde{i}\tilde{j}\gamma}
V_{\tilde{i}\tilde{j}\gamma\nu}(-\mathbf{q})
\sum_{\mathbf{k}}
\Biggl(
\sum_n
U_{\tilde{i}n}^*(\mathbf{k}-\mathbf{q})
A_{n\mathbf{k}-\mathbf{q}}^*
\Biggr)
\\
&\times
\Biggl(
\Sigma_{\tilde{i}\tilde{j}}^\gamma(\mathbf{k})
D_{\mathbf{k}}
\sum_m
U_{\tilde{j}m}(\mathbf{k})
A_{m\mathbf{k}}
\Biggr)
\\
&+
L_{\nu}(-\mathbf{q})
\sum_{\tilde{i}}\sum_{\mathbf{k}}
\Biggl(
\sum_n
U_{\tilde{i}n}^*(\mathbf{k}-\mathbf{q})
A_{n\mathbf{k}-\mathbf{q}}^*
\Biggr)
\Biggl(
D_{\mathbf{k}}
\sum_m
U_{\tilde{i}m}(\mathbf{k})
A_{m\mathbf{k}}
\Biggr).
\end{split}
\end{equation}

\begin{equation}
\begin{split}
2\operatorname{Re}\{B_{\nu\mathbf{q}}\}
\sum_{n\mathbf{k}}D_{\mathbf{k}+\mathbf{q}}
\sum_{m\nu'\mathbf{q}'}
g_{mn\nu'}(\mathbf{k},\mathbf{q}')
A_{m\mathbf{k}+\mathbf{q}'}^*
A_{n\mathbf{k}}
B_{\nu'-\mathbf{q}'}^*
={}&
2\operatorname{Re}\{B_{\nu\mathbf{q}}\}
\sum_{\mathbf{k}}D_{\mathbf{k}+\mathbf{q}}
\sum_{\gamma}\sum_{\tilde{i}\tilde{j}}
\Biggl(
\sum_n
U_{\tilde{j}n}(\mathbf{k})A_{n\mathbf{k}}
\Biggr)
\Sigma_{\tilde{i}\tilde{j}}^\gamma(\mathbf{k})
\\
&\times
\sum_{\mathbf{q}'}
\Biggl(
\sum_m
U_{\tilde{i}m}^*(\mathbf{k}+\mathbf{q}')
A_{m\mathbf{k}+\mathbf{q}'}^*
\Biggr)
\Biggl(
\sum_{\nu'}
V_{\tilde{i}\tilde{j}\gamma\nu'}(\mathbf{q}')
B_{\nu'-\mathbf{q}'}^*
\Biggr)
\\
&+
2\operatorname{Re}\{B_{\nu\mathbf{q}}\}
\sum_{\mathbf{k}}D_{\mathbf{k}+\mathbf{q}}
\sum_{\tilde{i}}
\Biggl(
\sum_n
U_{\tilde{i}n}(\mathbf{k})A_{n\mathbf{k}}
\Biggr)
\\
&\times
\sum_{\mathbf{q}'}
\Biggl(
\sum_{\nu'}
L_{\nu'}(\mathbf{q}')
B_{\nu'-\mathbf{q}'}^*
\Biggr)
\Biggl(
\sum_m
U_{\tilde{i}m}^*(\mathbf{k}+\mathbf{q}')
A_{m\mathbf{k}+\mathbf{q}'}^*
\Biggr).
\end{split}
\end{equation}
\end{subequations}

\subsection{Diagonal Elements of Hessian}
For approximate second-order methods such as the GDM, the diagonal elements of the Hessian are useful for preconditioning. In the followings, we list necessary equations for implementing the diagonal elements of the Hessian.
\begin{equation}
\begin{split}
\frac{\partial^2 E_{\mathrm{dD2}}}{\partial \operatorname{Re}\{A_{n\mathbf{k}}\}^2}
={}&
2\langle\Psi|\Psi\rangle^{-1}
\Biggl[
\Biggl(
\varepsilon_{n\mathbf{k}}
- E_{\mathrm{dD2}}(A,B)
- \operatorname{Re}\{A_{n\mathbf{k}}\}
\frac{\partial E_{\mathrm{dD2}}(A,B)}{\partial \operatorname{Re}\{A_{n\mathbf{k}}\}}
\Biggr)
D_{\mathbf{k}}
\\
&\qquad
+ \sum_{\nu\mathbf{q}}
\omega_{\nu\mathbf{q}} |B_{\nu\mathbf{q}}|^2 D_{\mathbf{k}+\mathbf{q}}
+ 2\operatorname{Re}\Biggl\{
\sum_{\nu'}
g_{nn\nu'}(\mathbf{k},0)\,
B_{\nu'\mathbf{0}}^*\,D_{\mathbf{k}}
\Biggr\}
\Biggr].
\end{split}
\end{equation}

\begin{equation}
\begin{split}
\frac{\partial^2 E_{\mathrm{dD2}}}{\partial \operatorname{Im}\{A_{n\mathbf{k}}\}^2}
={}&
2\langle\Psi|\Psi\rangle^{-1}
\Biggl[
\Biggl(
\varepsilon_{n\mathbf{k}}
- E_{\mathrm{dD2}}(A,B)
- \operatorname{Im}\{A_{n\mathbf{k}}\}
\frac{\partial E_{\mathrm{dD2}}(A,B)}{\partial \operatorname{Im}\{A_{n\mathbf{k}}\}}
\Biggr)
D_{\mathbf{k}}
\\
&\qquad
+ \sum_{\nu\mathbf{q}}
\omega_{\nu\mathbf{q}} |B_{\nu\mathbf{q}}|^2 D_{\mathbf{k}+\mathbf{q}}
+ 2\operatorname{Re}\Biggl\{
\sum_{\nu'}
g_{nn\nu'}(\mathbf{k},0)\,
B_{\nu'\mathbf{0}}^*\,D_{\mathbf{k}}
\Biggr\}
\Biggr].
\end{split}
\end{equation}

\begin{equation}
\begin{split}
\frac{\partial^2 E_{\mathrm{dD2}}}{\partial \operatorname{Re}\{B_{\nu\mathbf{q}}\}^2}
={}&
2\langle\Psi|\Psi\rangle^{-1}
\Biggl[
\sum_{n\mathbf{k}}
\Biggl(
\varepsilon_{n\mathbf{k}} + \omega_{\nu\mathbf{q}}
- E_{\mathrm{dD2}}(A,B)
- \operatorname{Re}\{B_{\nu\mathbf{q}}\}
\frac{\partial E_{\mathrm{dD2}}(A,B)}{\partial \operatorname{Re}\{B_{\nu\mathbf{q}}\}}
\Biggr)
|A_{n\mathbf{k}}|^2 D_{\mathbf{k}+\mathbf{q}}
\\
&\qquad
+ 2\operatorname{Re}\{B_{\nu\mathbf{q}}\}^2
\sum_{n\mathbf{k}}
\Bigl(
\varepsilon_{n\mathbf{k}} + 2\omega_{\nu\mathbf{q}} - E_{\mathrm{dD2}}(A,B)
\Bigr)
|A_{n\mathbf{k}}|^2 D_{\mathbf{k}+2\mathbf{q}}
\\
&\qquad
+ \sum_{n\mathbf{k}} |A_{n\mathbf{k}}|^2
\sum_{\nu'\mathbf{q}'}
\omega_{\nu'\mathbf{q}'} |B_{\nu'\mathbf{q}'}|^2
D_{\mathbf{q}'+\mathbf{k}+\mathbf{q}}
\\
&\qquad
+ 2\operatorname{Re}\{B_{\nu\mathbf{q}}\}^2
\sum_{n\mathbf{k}} |A_{n\mathbf{k}}|^2
\sum_{\nu'\mathbf{q}'}
\omega_{\nu'\mathbf{q}'} |B_{\nu'\mathbf{q}'}|^2
D_{\mathbf{k}+\mathbf{q}'+2\mathbf{q}}
\\
&\qquad
+ \operatorname{Re}\Biggl\{
4\operatorname{Re}\{B_{\nu\mathbf{q}}\}
\Biggl(
\sum_{nm\mathbf{k}}
g_{nm\nu}(\mathbf{k},-\mathbf{q})
A_{n\mathbf{k}-\mathbf{q}}^*
A_{m\mathbf{k}}
D_{\mathbf{k}+\mathbf{q}}
\Biggr)
\\
&\qquad\qquad
+ 4\operatorname{Re}\{B_{\nu\mathbf{q}}\}^2
\sum_{\substack{\mathbf{k}\mathbf{q}'\\nm\nu'}}
g_{nm\nu'}(\mathbf{k},\mathbf{q}')
A_{n\mathbf{k}+\mathbf{q}'}^*
A_{m\mathbf{k}}
B_{\nu',-\mathbf{q}'}^*
D_{\mathbf{k}+2\mathbf{q}}
\\
&\qquad\qquad
+ 2
\sum_{\substack{\mathbf{k}\mathbf{q}'\\nm\nu'}}
g_{nm\nu'}(\mathbf{k},\mathbf{q}')
A_{n\mathbf{k}+\mathbf{q}'}^*
A_{m\mathbf{k}}
B_{\nu',-\mathbf{q}'}^*
D_{\mathbf{k}+\mathbf{q}}
\Biggr\}
\Biggr].
\end{split}
\end{equation}

\begin{equation}
\begin{split}
\frac{\partial^2 E_{\mathrm{dD2}}}{\partial \operatorname{Im}\{B_{\nu\mathbf{q}}\}^2}
={}&
2\langle\Psi|\Psi\rangle^{-1}
\Biggl[
\sum_{n\mathbf{k}}
\Biggl(
\varepsilon_{n\mathbf{k}} + \omega_{\nu\mathbf{q}}
- E_{\mathrm{dD2}}(A,B)
- \operatorname{Im}\{B_{\nu\mathbf{q}}\}
\frac{\partial E_{\mathrm{dD2}}(A,B)}{\partial \operatorname{Im}\{B_{\nu\mathbf{q}}\}}
\Biggr)
|A_{n\mathbf{k}}|^2 D_{\mathbf{k}+\mathbf{q}}
\\
&\qquad
+ 2\operatorname{Im}\{B_{\nu\mathbf{q}}\}^2
\sum_{n\mathbf{k}}
\Bigl(
\varepsilon_{n\mathbf{k}} + 2\omega_{\nu\mathbf{q}} - E_{\mathrm{dD2}}(A,B)
\Bigr)
|A_{n\mathbf{k}}|^2 D_{\mathbf{k}+2\mathbf{q}}
\\
&\qquad
+ \sum_{n\mathbf{k}} |A_{n\mathbf{k}}|^2
\sum_{\nu'\mathbf{q}'}
\omega_{\nu'\mathbf{q}'} |B_{\nu'\mathbf{q}'}|^2
D_{\mathbf{q}'+\mathbf{k}+\mathbf{q}}
\\
&\qquad
+ 2\operatorname{Im}\{B_{\nu\mathbf{q}}\}^2
\sum_{n\mathbf{k}} |A_{n\mathbf{k}}|^2
\sum_{\nu'\mathbf{q}'}
\omega_{\nu'\mathbf{q}'} |B_{\nu'\mathbf{q}'}|^2
D_{\mathbf{k}+\mathbf{q}'+2\mathbf{q}}
\\
&\qquad
+ \operatorname{Re}\Biggl\{
4\operatorname{Im}\{B_{\nu\mathbf{q}}\}
\Biggl(
\sum_{nm\mathbf{k}}
g_{nm\nu}(\mathbf{k},-\mathbf{q})
A_{n\mathbf{k}-\mathbf{q}}^*
A_{m\mathbf{k}}
D_{\mathbf{k}+\mathbf{q}}
\Biggr)
\\
&\qquad\qquad
+ 4\operatorname{Im}\{B_{\nu\mathbf{q}}\}^2
\sum_{\substack{\mathbf{k}\mathbf{q}'\\nm\nu'}}
g_{nm\nu'}(\mathbf{k},\mathbf{q}')
A_{n\mathbf{k}+\mathbf{q}'}^*
A_{m\mathbf{k}}
B_{\nu',-\mathbf{q}'}^*
D_{\mathbf{k}+2\mathbf{q}}
\\
&\qquad\qquad
+ 2
\sum_{\substack{\mathbf{k}\mathbf{q}'\\nm\nu'}}
g_{nm\nu'}(\mathbf{k},\mathbf{q}')
A_{n\mathbf{k}+\mathbf{q}'}^*
A_{m\mathbf{k}}
B_{\nu',-\mathbf{q}'}^*
D_{\mathbf{k}+\mathbf{q}}
\Biggr\}
\Biggr].
\end{split}
\end{equation}

Again, for the decomposition of the electron--phonon matrix elements, one needs to adjust the contractions accordingly. For the diagonal elements of the Hessian we find that an approximation is needed to retain the linear scaling in $N_{\mathbf{k}}$. We find that averaging $D_{\mathbf{k}+\mathbf{q}}$ leads to a stable convergence, especially for weakly coupled systems. 

\begin{subequations}
\begin{equation}
\begin{split}
4\operatorname{Re}\{B_{\nu \mathbf{q}}\}
&\sum_{nm\mathbf{k}}g_{nm\nu}(\mathbf{k},-\mathbf{q})
A_{n\mathbf{k}-\mathbf{q}}^{*} A_{m\mathbf{k}} D_{\mathbf{k}+\mathbf{q}}
\\
&=
4\operatorname{Re}\{B_{\nu \mathbf{q}}\}
\sum_{\mathbf{k}\tilde{i}\tilde{j}\gamma}
\left( \sum_n U_{\tilde{i}n}^*({\mathbf{k}-\mathbf{q}})A_{n\mathbf{k}-\mathbf{q}}^{*} \right)
\left( \sum_m U_{\tilde{j}m}({\mathbf{k}})A_{m\mathbf{k}} \right)
V_{\tilde{i}\tilde{j}\gamma\nu}(-\mathbf{q})
\Sigma_{\tilde{i}\tilde{j}\gamma}(\mathbf{k})
\underbrace{D_{\mathbf{k}+\mathbf{q}}}_{\approx \langle D\rangle}
\\
&\quad
+ 4\operatorname{Re}\{B_{\nu \mathbf{q}}\} L_{\nu}(-\mathbf{q})
\sum_{\mathbf{k}\tilde{i}}
\left( \sum_n U_{\tilde{i}n}^{*}(\mathbf{k}-\mathbf{q}) A_{n\mathbf{k}-\mathbf{q}}^{*}\right)
\left(\sum_m U_{\tilde{i}m}(\mathbf{k}) A_{m\mathbf{k}}\right)
\underbrace{D_{\mathbf{k}+\mathbf{q}}}_{\approx \langle D\rangle}
\\
&\approx
4\operatorname{Re}\{B_{\nu \mathbf{q}}\} \langle D\rangle
\sum_{\tilde{i}\tilde{j}\gamma}V_{\tilde{i}\tilde{j}\gamma\nu}({-\mathbf{q}} )
\sum_{\mathbf{k}}
\left( \Sigma_{\tilde{i}\tilde{j}\gamma}(\mathbf{k}) \sum_m U_{\tilde{j}m}({\mathbf{k}})A_{m\mathbf{k}} \right)
\left( \sum_n U_{\tilde{i}n}^{*}(\mathbf{k}-\mathbf{q})A_{n\mathbf{k}-\mathbf{q}}^{*}\right)
\\
&\quad
+ 4\operatorname{Re}\{B_{\nu \mathbf{q}}\} \langle D\rangle L_{\nu}(-\mathbf{q})
\sum_{\mathbf{k}\tilde{i}}
\left( \sum_n U_{\tilde{i}n}^{*}(\mathbf{k}-\mathbf{q}) A_{n\mathbf{k}-\mathbf{q}}^{*}\right)
\left(\sum_m U_{\tilde{i}m}(\mathbf{k}) A_{m\mathbf{k}}\right)
\end{split}
\end{equation}

\begin{equation}
\begin{split}
4\operatorname{Re}\{B_{\nu \mathbf{q}}\}^2
&\sum_{\mathbf{k}\mathbf{q}'nm\nu'} g_{nm\nu'}(\mathbf{k},\mathbf{q}')
A_{n\mathbf{k}+\mathbf{q}'}^{*}A_{m\mathbf{k}} B_{\nu'-\mathbf{q}'}^* D_{\mathbf{k}+2\mathbf{q}}
\\
&=
4\operatorname{Re}\{B_{\nu \mathbf{q}}\}^2\sum_{\mathbf{k}\tilde{j}\tilde{i}\gamma}
D_{\mathbf{k}+2\mathbf{q}}
\left( \sum_m U_{\tilde{j}m}({\mathbf{k}})A_{m\mathbf{k}}\right)
\Sigma_{\tilde{i}\tilde{j}\gamma}(\mathbf{k})
\\
&\quad\times
\sum_{\mathbf{q}'}
\left( \sum_n U_{\tilde{i}n}^{*}(\mathbf{k}+\mathbf{q}') A_{n\mathbf{k}+\mathbf{q}}^{*}\right)
\left( \sum_\mu B_{\nu'-\mathbf{q}'}^* V_{\tilde{i}\tilde{j}\gamma\nu'}(\mathbf{q}')\right)
\\
&\quad
+4\operatorname{Re}\{B_{\nu \mathbf{q}}\}^2 \sum_\mathbf{k}
D_{\mathbf{k}+2\mathbf{q}}
\Biggl[
\sum_{\tilde{i}}
\left(\sum_m U_{\tilde{i}m}(\mathbf{k}) A_{m\mathbf{k}}\right)
\sum_{\mathbf{q}'}
\left( \sum_{\nu'} L_{\nu'}(\mathbf{q}') B_{\nu'-\mathbf{q}'}^* \right)
\\
&\qquad\qquad\qquad\qquad\times
\left(\sum_{n}U_{\tilde{i}n}^*({\mathbf{k}+\mathbf{q}'})A_{n\mathbf{k}+\mathbf{q}'}^{*} \right)
\Biggr]
\end{split}
\end{equation}

\begin{equation}
\begin{split}
\sum_{\mathbf{k}\mathbf{q}'nm\nu'}
&g_{nm\nu'}(\mathbf{k},\mathbf{q}')
A_{n\mathbf{k}+\mathbf{q}'}^{*}A_{m\mathbf{k}} B_{\nu'-\mathbf{q}'}^* D_{\mathbf{k}+\mathbf{q}}
\\
&=
\sum_{\mathbf{k}}
\bigg(\sum_{m}U_{\tilde{j}m}(\mathbf{k}) A_{m\mathbf{k}}\bigg)
D_{\mathbf{k}+\mathbf{q}} \Sigma_{\tilde{i}\tilde{j}\gamma}({\mathbf{k}})
\\
&\quad\times
\sum_{\mathbf{q}'}
\left( \sum_n U_{\tilde{i}n}^*({\mathbf{k}+\mathbf{q}}) A_{n\mathbf{k}+\mathbf{q}}^{*}\right)
\left(\sum_{\nu'} V_{\tilde{i}\tilde{j}\gamma\nu'}(\mathbf{q}') B_{\nu'-\mathbf{q}'}^* \right)
\\
&\quad
+ \sum_{\mathbf{k}\tilde{i}}
D_{\mathbf{k}+\mathbf{q}}
\left(\sum_m U_{\tilde{i}m}({\mathbf{k}}) A_{m\mathbf{k}}\right)
\sum_{\mathbf{q}'}
\left(\sum_{\nu'} L_{\nu'}(\mathbf{q}')B_{\nu'-\mathbf{q}}^* \right)
\\
&\qquad\times
\left( \sum_n U_{\tilde{i}n}^*({\mathbf{k}+\mathbf{q}'}) A_{n\mathbf{k}+\mathbf{q}}^{*}\right)
\end{split}
\end{equation}

\begin{equation}
\begin{split}
4 \operatorname{Re}\left\{\sum_{\nu'} g_{nn\nu'}(\mathbf{k},0) B_{\nu'0}^* D_\mathbf{k}\right\}
&=
D_\mathbf{k}
\sum_{\tilde{i}\tilde{j}}
U_{\tilde{i}n}^*({\mathbf{k}}) U_{\tilde{j}n}(\mathbf{k})
\sum_\gamma \Sigma_{\tilde{i}\tilde{j}\gamma}(\mathbf{k})
\left( \sum_{\nu'} V_{\tilde{i}\tilde{j}\gamma\nu'}(0) B_{\nu'0}^*\right)
\\
&\quad
+ \sum_{\nu} L_{\nu}(0) B_{\nu0}^*
\left(\sum_{\tilde{i}} U_{\tilde{i}n}^*({\mathbf{k}}) U_{\tilde{i}n}({\mathbf{k}})  \right)
\end{split}
\end{equation}
\end{subequations}

\section{Observables} \label{sec:observables}
\subsection{Phonon Occupations}\label{subsec:phonon_number}
\begin{figure}
    \centering
    \includegraphics[width=\linewidth]{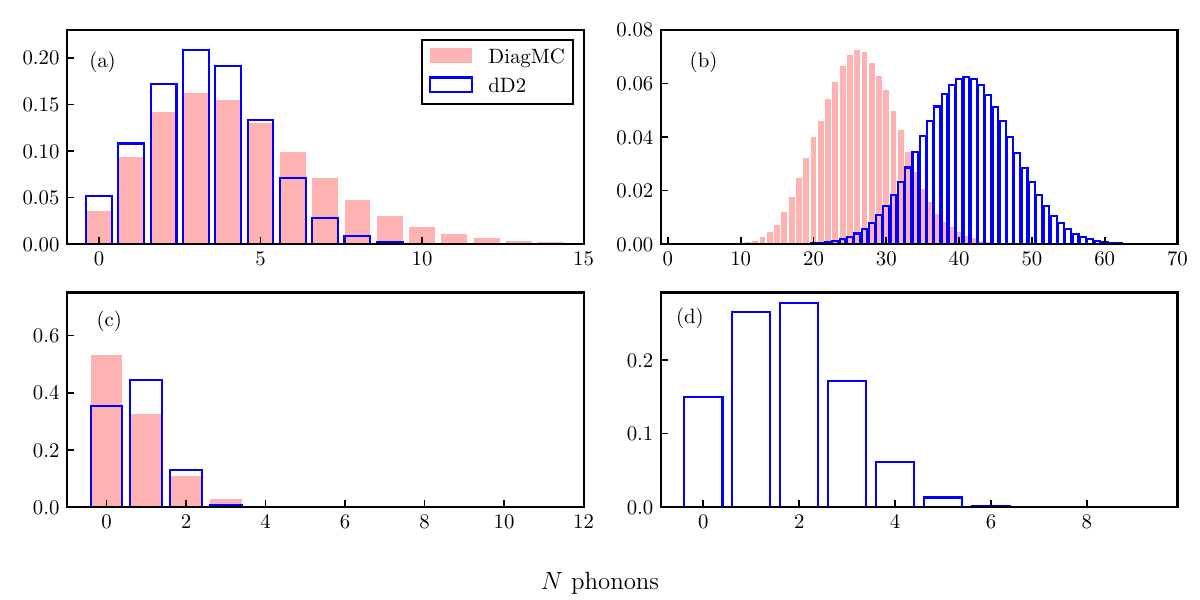}
    \caption{Phonon statistics for the (a) electron and (b) hole polaron in \ce{LiF}, and the electron polarons in (c) anatase and (d) rutile. dD2 bars are obtained from a Gaussian distribution with mean $\langle N_\text{ph}\rangle$ and standard deviation $\sqrt{\langle N_\text{ph}^2\rangle - \langle N_\text{ph}\rangle^2}$}
    \label{fig:phonons}
\end{figure}

We compute the average number of phonons $\langle N_{\text{ph}}\rangle$ in the ground state as well as the second central moment of the corresponding probability distribution:
\begin{align}
\begin{split}
    \big\langle N_{\text{ph}}\big\rangle &= \langle\Psi|\Psi\rangle^{-1} \left(\sum_n \mathrm{e}^{\mathrm{i}\mathbf{K} \cdot\mathbf{R}_n} \mathrm{e}^{\sum_{\nu\mathbf{q}}|B_{\nu \mathbf{q}}|^2 \mathrm{e}^{-\mathrm{i}\mathbf{q}\cdot\mathbf{R}_n}} \left[\sum_{j\mathbf{k}}|A_{j\mathbf{k}}|^2 \mathrm{e}^{-\mathrm{i}\mathbf{k}\cdot\mathbf{R}_n}\right] \left[\sum_{\nu'\mathbf{q}'}|B_{\nu'\mathbf{q}'}|^2 \mathrm{e}^{-\mathrm{i}\mathbf{q}'\cdot\mathbf{R}_n}\right]\right)
\end{split}\\
\begin{split}
    \big\langle \big(N_{\text{ph}} - \big\langle N_{\text{ph}}\rangle\big)^2 \big\rangle &= \langle\Psi|\Psi\rangle^{-1} \left(\sum_n \mathrm{e}^{\mathrm{i}\mathbf{K}\cdot \mathbf{R}_n} \mathrm{e}^{\sum_{\nu\mathbf{q}}|B_{\nu \mathbf{q}}|^2 \mathrm{e}^{-\mathrm{i}\mathbf{q}\cdot\mathbf{R}_n}} \left[\sum_{j\mathbf{k}}|A_{j\mathbf{k}}|^2 \mathrm{e}^{-\mathrm{i}\mathbf{k}\cdot\mathbf{R}_n}\right] \left[\sum_{\nu'\mathbf{q}'}|B_{\nu'\mathbf{q}'}|^2 \mathrm{e}^{-\mathrm{i}\mathbf{q}'\cdot\mathbf{R}_n}\right]^2\right)\\
    &\qquad+ \langle N_{\mathrm{ph}}\rangle - \langle N_{\mathrm{ph}}\rangle^2    
\end{split}
\end{align}
Corresponding Gaussian distributions are shown in Fig.~\ref{fig:phonons}.

We can also evaluate the number of phonons expected in each mode for each $\mathbf{q}$-point,
\begin{align}
    \langle b_{\nu\mathbf{q}}^\dagger b_{\nu\mathbf{q}}\rangle &=  \langle \Psi|\Psi\rangle^{-1} |B_{\nu \mathbf{q}}|^2 \sum_{n\mathbf{k}} |A_{n\mathbf{k}}|^2 D_{\mathbf{k}+\mathbf{q}}.
\end{align}
Fig.~\ref{fig:elec_dens} shows this quantity on top of the phonon bands.

\subsection{Carrier Occupations} \label{subsec:elec_obs}
\begin{figure}
    \centering
    \includegraphics[width=\linewidth]{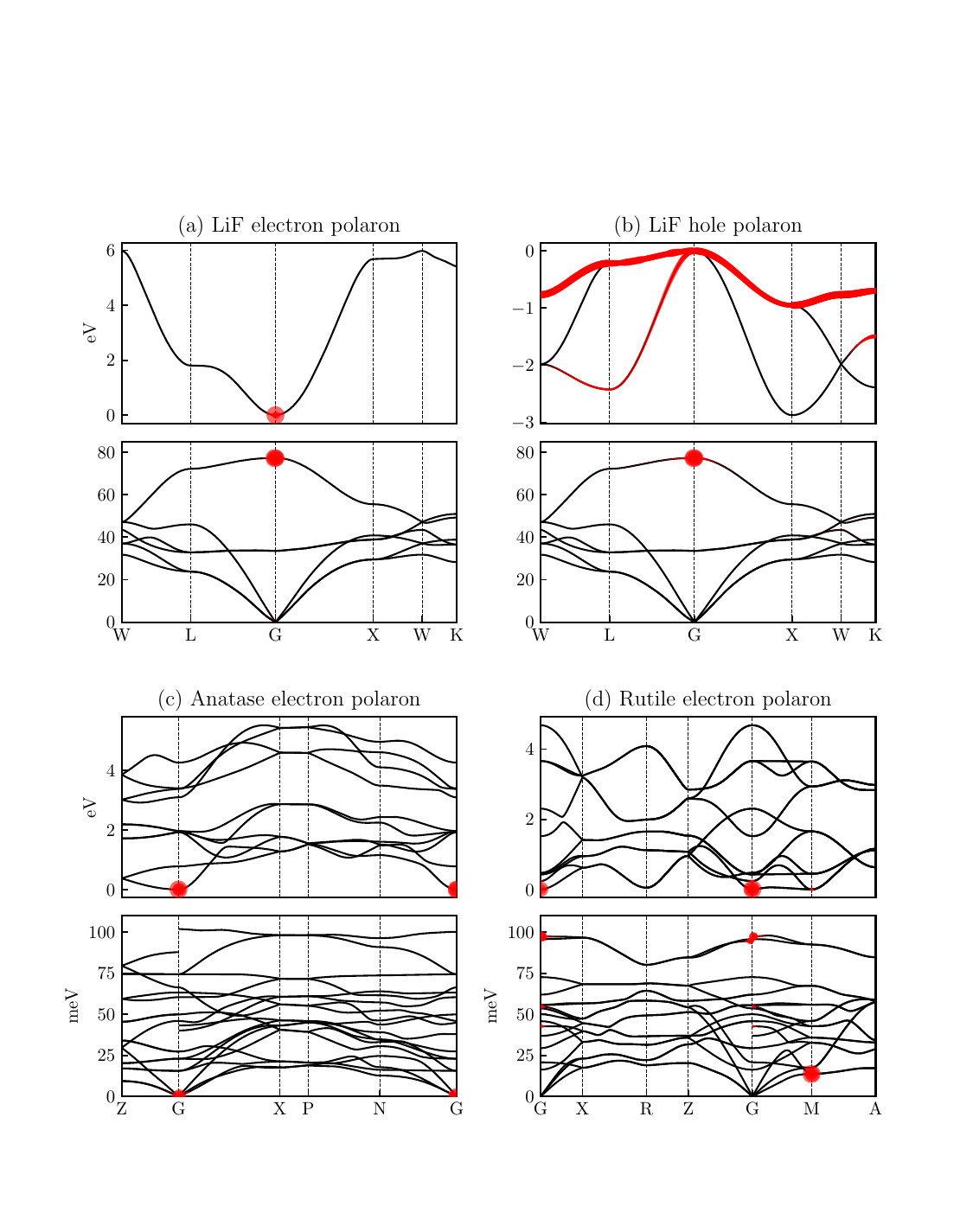}
    \caption{(a) Electron density $\langle c_{n\textbf{k}}^\dagger c_{n\textbf{k}}\rangle$ (upper panel) and phonon number $\langle b_{\nu\textbf{q}}^\dagger b_{\nu\textbf{q}}\rangle$ (lower panel) obtained by interpolation from dD2 wavefunction for the electron polaron in \ce{LiF}, converged on $101^3$ \textbf{k}-mesh, using cubic splines. Radius of red dots is proportional to the corresponding interpolated quantity at that $\mathbf{k}$-point. (b) Same observables as in (a) for the hole polaron in \ce{LiF}, and the electron polarons in (c) anatase and (d) rutile.}
    \label{fig:elec_dens}
\end{figure}

To further characterize the ground states obtained, we compute the momentum-resolved electronic densities,
\begin{align}
    \langle c_{n\mathbf{k}}^\dagger c_{n\mathbf{k}}\rangle = \langle \Psi|\Psi\rangle^{-1} \:|A_{n\mathbf{k}}|^2 D_{\mathbf{k}}.
\end{align}
Since our $\mathbf{k}$-mesh is uniformly discretized, we use cubic splines to subsample along the high-symmetry points within the unit cell. Electronic densities are shown in Fig.~\ref{fig:elec_dens}. 

\subsection{Polaron Extent} \label{sec:polaron_extent}
In order to make a prediction on the polaron size of a system we investigate the density-displacement correlation
\begin{align}
    d_{n,\kappa \alpha}(\mathbf{R}_\text{p}) = \left \langle \sum_{\mathbf{R}_\text{e}} \hat{n}_{n \mathbf{R}_\text{e}} \hat{u}^{\kappa \alpha}_{\mathbf{R}_\text{e} + \mathbf{R}_\text{p}} \right\rangle,
\end{align}
where 
\begin{align}
    \hat{n}_{n\mathbf{R}_\text{e}} &= c_{n \mathbf{R}_\text{e}}^\dagger c_{n \mathbf{R}_\text{e}} = \frac{1}{N_\mathbf{k}} \sum_{\mathbf{k}_1 \mathbf{k}_2 j_1 j_2} \mathrm{e}^{-\mathrm{i}\mathbf{R}_{\text{e}} \cdot(\mathbf{k}_1 - \mathbf{k_2})} U_{j_1 n}^*({\mathbf{k}_1}) U_{j_2 n}({\mathbf{k}_2}) c_{j_1 \mathbf{k}_1}^\dagger c_{j_2 \mathbf{k}_2} \\
    \hat{u}^{\kappa \alpha}_{\mathbf{R}_\text{e} + \mathbf{R}_\text{p}} &= \sum_{\nu\mathbf{q}} \sqrt{\frac{\hbar}{2 M_\kappa \omega_{\nu \mathbf{q}}}} e_{\kappa \alpha}^\nu(\mathbf{q}) \frac{\mathrm{e}^{\mathrm{i}\mathbf{q}\cdot(\mathbf{R}_\text{e} + \mathbf{R}_\text{p})}}{\sqrt{N_k}} \left( b_{\nu\mathbf{q}} + b_{\nu-\mathbf{q}}^\dagger \right).
\end{align}
We can sum out $\mathbf{R}_{\text{e}}$, and arrive at 
\begin{align}
    \sum_{\mathbf{R}_{\text{e}}}  \hat{n}_{n\mathbf{R}_\text{e}}\hat{u}^{\kappa \alpha}_{\mathbf{R}_\text{e} + \mathbf{R}_\text{p}} = \frac{1}{\sqrt{N_\mathbf{k}}}\sum_{\mathbf{k}\mathbf{q}ij\nu} U_{in}^*(\mathbf{k}+\mathbf{q})\: U_{jn}({\mathbf{k}})\: \underbrace{\sqrt{\frac{\hbar }{2 M_\kappa \omega_{\nu\mathbf{q}}}} e_{\kappa\alpha}^{\nu}(\mathbf{q}) }_{C_\nu^{\kappa \alpha} (\mathbf{q})}\: \mathrm{e}^{\mathrm{i}\mathbf{q}\cdot\mathbf{R}_\text{p}}\: c_{i\mathbf{k}+\mathbf{q}}^\dagger c_{j\mathbf{k}} \left(b_{\nu\mathbf{q}} + b_{\nu -\mathbf{q}}^\dagger \right).
\end{align}
For the dD2 expectation value we find similar to the electron--phonon contribution to the energy expectation
\begin{align}
    \left\langle \sum_{\mathbf{R}_{\text{e}}}  \hat{n}_{n\mathbf{R}_\text{e}}\hat{u}^{\kappa \alpha}_{\mathbf{R}_\text{e} + \mathbf{R}_\text{p}} \right \rangle = \frac{2}{\sqrt{N_\mathbf{k}}} \: \operatorname{Re}\left\{  \sum_{\nu\mathbf{q}} \mathrm{e}^{\mathrm{i}\mathbf{q}\cdot\mathbf{R}_\text{p}} C_{\nu}^{\kappa\alpha} (\mathbf{q}) B_{\nu -\mathbf{q}}^* \sum_\mathbf{k}D_\mathbf{k} \tilde{A}_{n\mathbf{k}}\tilde{A}_{n\mathbf{k}+\mathbf{q}}^*\right\}
\end{align}
exploiting TRS of the phonon modes, $e^\nu_{\kappa\alpha}(\mathbf{q}) = (e^\nu_{\kappa\alpha}(\mathbf{-q}))^*$, and frequencies, $\omega_{\nu\mathbf{q}} = \omega_{\nu-\mathbf{q}}$.

\begin{table}[h]
    \centering
    {\setlength{\tabcolsep}{8pt}
    \begin{tabular}{lcccc}
     \hline
     \hline
      &  LiF (h) &Rutile &  LiF (e)  & Anatase \\ [0.5ex]
     \hline
     $\text{FWHM}_1 \:/\: \si{\angstrom} $ & 6.77 & 25.17 &  46.19   & 64.06  \\
     $\text{FWHM}_2 \:/\: \si{\angstrom}$  & 3.56 &  21.94&  45.02 & 63.10 \\
     $\text{FWHM}_3 \:/\: \si{\angstrom}$  & 3.53 &  18.69 & 42.78 &  29.41  \\ 
     $R^2$  &0.91 & 0.67 & 0.83 & 0.78\\
     [0.5ex]
     \hline
     \hline
    \end{tabular}}
    \caption{FWHM along the principal axes of the covariance matrix $\Sigma$ obtained by fitting $|\sum_n \eta_{n\kappa\alpha}(\mathbf{R}_p)|_\parallel$ to a three-dimensional Gaussian.}
    \label{tab:fwhm}
\end{table}

In order to quantify the polaron extent we fit a 3-dimensional Gaussian
\begin{align} \
    u(\mathbf{r}) = A \exp\left(-\frac{1}{2} \left(\mathbf{r}-\mathbf{r}_0\right)\Sigma^{-1}\left(\mathbf{r}-\mathbf{r}_0\right)\right).
\end{align}
to the absolute value of the displacement correlation defined in \cref{eq:extent}. Here, $\Sigma$ is the positive semi-definite covariance matrix, parametrized by 6 independent parameters, $\mathbf{r}_0$ is the polaron center, and $A$ is the amplitude of the Gaussian. By diagonalizing the covariance matrix, one can identify the principal axes of the polaron,
\begin{align}
    u(\mathbf{r}) = A \exp\left(-\frac{1}{2} \sum_j \frac{y_j^2}{\lambda_j}\right),
\end{align}
where $\mathbf{y} = Q^T(\mathbf{r}-\mathbf{r}_0)$, and $\sum_j\Sigma_{ij}Q_{jl} = \lambda_{l}Q_{il}$. The exponent describes an ellipse, and after fitting all free parameters to the absolute displacement correlation, we find the optimal ellipse encapsulating the correlator. 
While a Gaussian fit is most effective in the strong coupling limit, the resulting FWHM values may well be used to obtain a qualitative measure for the real-space extent of larger polarons. 
The result of the fit and FWHM is given in \cref{tab:fwhm}.

\begin{figure*}
    \centering
    \includegraphics[width=\linewidth]{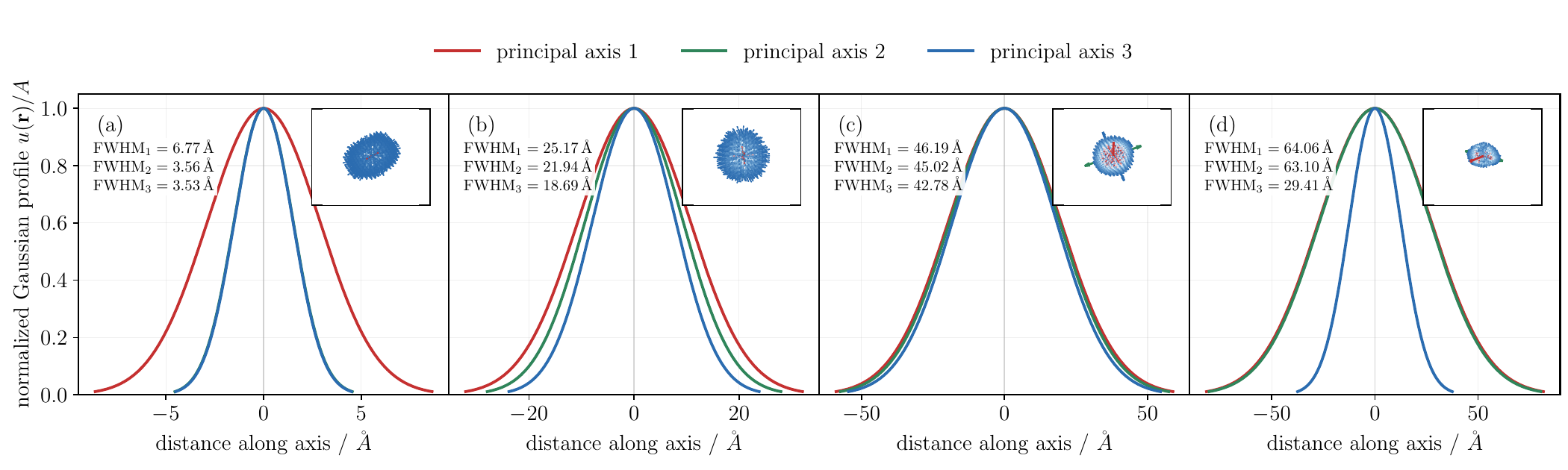}
    \caption{Polaron spatial extent for (a) \ce{LiF} hole polaron, and electron polarons in (b) rutile, (c) \ce{LiF}, and (d) anatase. Insets show directionality of the principle axes in the supercell. Shapes of the polaron are obtained from $\sum_n \eta_{n\kappa\alpha}(\mathbf{R}_j)$, and shown in (a) for the displacement correlation of F atoms, (b) Li atoms, and (c) and (d) for O atoms.}
    \label{fig:3d_extent}
\end{figure*}

\section{Additional computational details} \label{sec:comp_details}
To variationally optimize D2 and dD2 wavefunction parameters, we use geometric direct minimization (GDM)~\cite{van_voorhis_geometric_2002} as implemented in \texttt{Q-Chem}~\cite{epifanovsky_software_2021}. GDM is a second-order global optimization algorithm that relies on renormalizing degrees of freedom in the variational space. By preconditioning with the diagonal elements of the Hessian, disparate timescales can be treated on an equal footing, which we found critical for avoiding flat patches where energy varies little because gradients are dominated by bosonic parameter variation. 
We provide gradients and diagonal elements of the Hessian for the dD2 wavefunction in Sec.~\ref{sec:derivatives}.

To obtain Kohn--Sham bands and phonon frequencies, we run DFT and DFPT calculations using \texttt{Quantum Espresso}~\cite{giannozzi_quantum_2009}. For \ce{LiF} and anatase we use norm-conserving PBE pseudopotentials~\cite{schlipf_optimization_2015,hamann_optimized_2013}, and a plane wave cutoff of 150~\si{\rydberg} and 90~\si{\rydberg} respectively. 
Subsequently, we used \texttt{Wannier90}~\cite{pizzi_wannier90_2020} to obtain Wannier functions, allowing for an effective Wannier interpolation to ultimately obtain matrix elements at fine Brillouin-zone grids using an in-house version of \texttt{PERTURBO}~\cite{zhou_perturbo_2021}.
We used a uniform $12\times12\times12$ Brillouin-zone grid for DFT and DFPT calculation for LiF and a $6\times6\times6$ Brillouin-zone grid for corresponding calculations for anatase. 
 
Rutile matrix elements were obtained using intermediate files used by Luo \textit{et al.}~\cite{luo_first-principles_2025} They were initially generated by DFT+U and DFPT+U calculations on a coarse $4\times4\times6$ $\mathbf{k}$-mesh, using a Hubbard-U of 4.5~\si{\electronvolt}, and PBEsol ultrasoft pseudo-potentials.

\end{document}